\documentclass[letterpaper, twocolumn, superscriptaddress, showpacs, preprintnumbers, amsmath, amssymb,longbibliography, nofootinbib]{revtex4-1}

\usepackage{graphicx}
\usepackage{dcolumn}
\usepackage{bm}
\usepackage{color}
\usepackage{hyperref}
\usepackage{caption}
\usepackage{adjustbox}
\usepackage[lofdepth,lotdepth]{subfig}
\usepackage{caption}

\def\be{\begin{equation}}
\def\ee{\end{equation}}
\def\bea{\begin{eqnarray}}
\def\eea{\end{eqnarray}}
\def\bfq{\mathbf{q}}
\def\bfp{\mathbf{p}}
\def\bfA{\mathbf{A}}
\def\bfa{\mathbf{a}}
\def\bfr{\mathbf{r}}
\def\bfD{\mathbf{D}}

\def\bfe{\mathbf{e}}
\def\bfj{\mathbf{j}}
\def\bfk{\mathbf{k}}
\def\bfF{\mathbf{F}}
\def\bfv{\mathbf{v}}

\def\cf{{\mbox{\tiny{cf}}}}
\def\m{m^*}

\def\cs{{\mbox{\tiny{CS}}}}
\def\df{{\mbox{\tiny{DF}}}}

\def\nn{\nonumber\\}

\begin{document}

\title{Particle-Hole Symmetry in the Fermion-Chern-Simons and Dirac Descriptions of a Half-Filled Landau Level}

\author {Chong Wang}
\affiliation{Department of Physics, Harvard University, Cambridge MA 02138, USA}

\author{Nigel R. Cooper}
\affiliation{T.C.M. Group, Cavendish Laboratory, University of Cambridge,
JJ Thomson Avenue, Cambridge, CB3 0HE, U.K.} 

\author{Bertrand I. Halperin} 
\affiliation{Department of Physics, Harvard University, Cambridge MA 02138, USA}

\author{Ady Stern} 
\affiliation{ Department of Condensed Matter Physics, Weizmann Institute of Science, Rehovot, Israel 76100  }

\begin{abstract}

It is well known that there is a particle-hole symmetry for spin-polarized electrons with two-body interactions in a partially filled Landau level, which becomes exact in the limit where the cyclotron energy is large compared to the interaction strength, so one can ignore mixing between Landau levels. This symmetry is explicit in the description of a half-filled Landau level recently introduced by  D. T. Son, using Dirac fermions, but it was thought to be absent in the older fermion-Chern-Simons approach, developed by Halperin, Lee, and Read and subsequent authors. We show here, however, that when properly evaluated, the Halperin, Lee, Read (HLR) theory gives results for long-wavelength low-energy physical properties, including the Hall conductance in the presence of impurities and the  positions of minima in the magnetoroton spectra for fractional quantized Hall states close to half-filling, that are identical to predictions of the Dirac formulation.  In fact,  the HLR theory predicts an emergent particle-hole symmetry near half filling, even when the cyclotron energy is finite.

\end{abstract}
\maketitle

\today

 \tableofcontents

\section{Introduction}
\label{sec:intro}

A series of recent developments have focused renewed attention on the problem of a two-dimensional system of interacting electrons at, or close to, a half-filled Landau level. In particular, in a highly original work, D. T. Son\cite{sonphcfl} has proposed a description of the  half-filled Landau level that employs a collection of relativistic Dirac fermions, interacting with an emergent gauge field with no Chern-Simons term. This description stands in contrast to  the more traditional description in terms of non-relativistic ``composite fermions" interacting with a Chern-Simons gauge field,  developed by Halperin, Lee and Read (HLR)\cite{hlr} and others,  some twenty years ago. (See, e.g., Refs.~\cite{jaincf,kalmeyer92,lopez91,greiter92,rejaei92,kim94,altshuler94,SimonHalperin,sternSH95}).

The Son-Dirac description has led to a number of valuable insights into the conventional problem of two-dimensional electrons in a strong magnetic field\cite{dualdrmaxav,cfltislrev,geraedtsnum,msgmp15,WS16}, and it has also served to elucidate connections to other physical problems, such as exotic electronic states that could arise at the surface of a three-dimensional topological insulator\cite{mrosscdl15,dualdrcwts2015,dualdrmaxav}, time-reversal invariant $U(1)$ quantum spin liquids in three dimensions\cite{tsymmu1,dualdrmaxav,bulkdualmax}, and a class of field theory dualities in $(2+1)$ dimensions\cite{dualdrcwts2015,dualdrmaxav,dualdrMAM,SSWW,KarchTong,Murugan,kachrubosonization}.

The Dirac picture seems to have
some significant advantages compared with the HLR description for the  conventional two-dimensional electron system,  in particular with respect to particle-hole (PH) symmetry. 
It  is well known that a partially-filled Landau level  of spin-polarized electrons  with two-body interactions should have an  exact PH  symmetry about half-filling, in the limit where the electron-electron interaction is weak compared to the cyclotron energy, so one can neglect mixing between Landau levels\cite{girvin84}. Numerical calculations, either through trial wave functions motivated by the composite fermi liquid picture\cite{rzyhald2000,balramjain}, or through unbiased energetic calculations\cite{rzyhald2000,geraedtsnum}, seem to confirm that this symmetry is unbroken in the incompressible phase. This symmetry is made manifest in the Dirac model by setting a single parameter equal to zero, the Dirac mass $m_D$.


By contrast, the HLR approach is not explicitly PH symmetric, and in fact it has  been questioned whether the  approach is even compatible with PH symmetry\cite{klkgphhlr,Barkeshli15}.
It has  been suggested that the Dirac theory and the HLR theory actually represent different fixed points and that there might necessarily be some kind of discontinuous phase transition separating these fixed points\cite{klkgphhlr,sonphcfl,Barkeshli15,PSV}.
These suggestions have been based on analyses of several key physical properties, in which  it appeared that predictions of HLR were contradictory to PH symmetry. 

In this paper, we reexamine several of these properties, and we find that 
when properly analyzed, the HLR theory gives identical results to the Dirac theory,  in the limit of  long wavelengths and  low energies,
near half filling.  Some of the confusion about these points has arisen simply because the predictions of HLR theory were not previously analyzed with sufficient care. 
Despite our limitations to long-wavelengths and low-energies,  we believe that our analysis casts strong doubt on the possibility that there is any regime of parameters in which the Dirac description and the HLR description correspond to two different phases of matter.
Specifically, we  have carried out detailed studies of two types of properties where it has been suggested that there are irreconcilable differences between the HLR and Son-Dirac descriptions -- the Hall conductance of a half-filled Landau level in the presence of disorder, and the momentum values of the minima in the magnetoroton spectra of fractional quantized Hall states that are symmetrically displaced from $\nu=1/2$.

In the presence of a
disorder potential that is statistically PH symmetric,  symmetry dictates that the Hall conductance should be exactly $e^2/2h$, in the absence of Landau level mixing.  Since 1997, it has been widely believed  that HLR is incompatible with this requirement, and that HLR implies deviations in the Hall conductance  proportional to the inverse square of the mean-free path of the composite fermions.
We show below, however, that when properly evaluated,  these deviations are absent in the HLR theory, at least in the case of weak, long-wavelength, disorder potentials.

For a system where the electronic filling factor $\nu$ is close to one half, oscillations in the conductivity at finite wave vector $q$ and frequency $\omega$ have been predicted, and in some cases observed, as a function of the deviations from half filling.  These oscillations involve excitation or modulation at a non-zero  wave vector  $q$, where maxima or minima in some characteristic of the response are predicted to occur at a series of values of $q$, {\em approximately} given by
\be
\label{qn1}
q_n \approx \frac{z_n |e \Delta B|}{\hbar k_F}
\ee
where $z_n$ is the $n$-th zero of the $J_1$ Bessel function, $\Delta B$ is the deviation of the magnetic field from the field at half filling, and $k_F$ is the Fermi wave vector of the composite fermions.  PH symmetry requires that if the electron density is varied, while the magnetic field is held fixed, the wave vectors $q_n$ should be precisely independent of the sign of $\Delta B$.  In the Son-Dirac theory,
Eq.~(\ref{qn1})  directly obeys this PH symmetry,  because  the value of $k_F$ is a constant, determined by the magnetic field,  independent of the electron density.  In HLR, however, $k_F$  is determined by the electron density, which will be slightly different for positive  and negative values of $\Delta B$.
Therefore, if one were to treat Eq.~(\ref{qn1}) as an exact equality, using the definition of $k_F$ in HLR, one would find that PH symmetry is obeyed to first order in $\Delta B$, but is violated at second order.

 We show below that a careful evaluation of the locations of minima in the magnetoroton excitation spectrum  in fractional quantum Hall states close to $\nu=1/2$, originally discussed by Simon and Halperin (SH)\cite{SimonHalperin}, using the HLR approach, gives
predictions that are  PH symmetric, at least to order $(\Delta B)^2$. We show that these predictions coincide with the predictions of the Son-Dirac theory. The SH formulas actually contain corrections to Eq.~(\ref{qn1}), which vanish in the limit $\Delta B \to 0$ but are non-zero at order $(\Delta B)^2$ and which  precisely eliminate the PH asymmetry  at this order.

We note that the results described above were both obtained by careful evaluation of the HLR theory at the RPA level, and did not require any explicit assumption of particle hole symmetry, or any apparent assumption about the ratio between  the  electron interaction strength and the bare electron cyclotron energy. These results suggest that even when this ratio is finite, so the electrons are {\em not} projected into a single Landau level, there may be an {\em emergent}  PH symmetry, {which becomes asymptotically exact in the limit of low-frequency, long-wavelength and small deviation from half-filling. Our results show that for the properties we have analyzed, this is true at least to some nontrivial orders in frequency, momentum and deviation from half-filling.}

Within the context of HLR theory, we find that a similar degree of PH symmetry should emerge in the vicinity of other fractions of the form $1/(2n)$, such as 1/4, 1/6,  etc.  As a practical matter, this is only of interest for small values of $n$, since at least for the case of Coulomb interactions between the electrons, the ground state  for values of $n>3$  appears to be a Wigner crystal of electrons, rather than a liquid of composite fermions. Nevertheless, an  emergent PH symmetry at $\nu = 1/4$ or $1/6$ would be noteworthy, since there is no exact particle hole symmetry about fractions other than 1/2, even for electrons confined to a single Landau level.

The structure of the paper is the following. In the next Section we review the HLR approach to the half-filled Landau level. In Section \ref{sec:dc} we address the issue of $dc$ transport at $\nu=1/2$ in the presence of disorder, and show how the HLR approach yields results which are consistent with the requirements of particle-hole symmetry. In Section \ref{sec:commens} 
we address ``commensurability oscillations", which occur at fillings slightly away from $\nu=1/2$, with a focus on the locations of minima in the dispersion curves for the lowest-energy magnetoroton excitations in fractional quantized Hall states near half filling. We show 
how an analysis  within the HLR approach yields results that are consistent with the requirements of particle-hole symmetry. In Section \ref{sec:dirac}, we review the Son-Dirac approach, and make a comparison between results of that approach and our analyses based on  HLR.  
We conclude with a Summary section. 

\section{Review of the HLR approach\label{sec:hlrreview}}

\subsection{Definition of the Problem \label{sec:problem}}

We consider a two-dimensional  system of interacting electrons in a strong magnetic field, with a Landau level filling fraction $\nu$ that is equal to or close to $\nu=1/2$.  We assume that the electrons are fully spin polarized, so we may neglect the spin degree of freedom. The Hamiltonian of the system may be written in the form
\be
H_0 = \sum_j \frac{|\bfp_j - \bfA (\bfr_j)|^2 }{2m  } +V_2,
\ee
where $V_2$ is a two-body interaction of the form
\be
V_2 =   \frac{1}{2} \sum_{i \neq j} v_2( \bfr_i - \bfr_j),
\ee
and $\bfA$ is the vector potential due to a uniform magnetic field $B$ in the $z$-direction.
In the case where $v_2$ is a long-range potential, the Hamiltonian must include interactions between the electrons and a uniform neutralizing background,  which we include in $V_2$.
In the presence of impurities, we  shall  add   a one-body potential $V_1(\bfr_j)$ which depends on position; for the present, however, we shall consider a system without impurities, so we take $V_1 = 0$.
Except where otherwise stated, we use units where the electron charge is positive and equal to unity, and $\hbar=c=1$.

The system under consideration has several important properties. First, it is Galilean invariant, so that it must obey Kohn's theorem, which states that the response to a uniform time-varying electric field should be the same as for a system of non-interacting electrons in the given magnetic field. Second, as mentioned in the Introduction,  in the limit where the electron mass $m$ is taken to zero, so that the cyclotron energy becomes infinite while the electron-electron interaction is held fixed, the system should manifest an exact PH symmetry about Landau-level filling fraction $\nu=1/2$.  We shall see to what extent these properties are preserved by  approximations that have been proposed for treating the system.

\subsection{The HLR hypothesis}

The fermion-Chern-Simons approach employed in HLR began with an exact unitary transformation, a singular gauge transformation, where the many-body electron wave function is multiplied by a phase factor that  depends on the positions of all the electrons, such that the transformed Hamiltonian acquires a Chern-Simons gauge field $a_\mu$, with $-2$  flux quanta attached to every electron.  The transformed problem may be expressed in Lagrangian form by the following Lagrangian density:
\be
\label{Lhlr}
\mathcal{L}_0  =  \bar{\psi}  \left( i D_t - \mu  +\frac {\bfD \cdot \bfD }{2m}
\right) \psi  -\frac {ada} {8 \pi }  + \mathcal{L}_{\rm {int}}
\ee
\be
a da \equiv  \varepsilon^{\mu \nu \lambda} a_\mu \partial_\nu a_\lambda
\ee
\be
D_\mu \equiv \partial_\mu + i \, (a_\mu + A_\mu) .
\ee
Taking the variation of the Lagrangian with respect to $a_0$, we obtain the constraint
\be
\nabla \times \bfa = - 4 \pi  \, \bar{\psi} \psi = - 4 \pi \, n_{\rm{el}}(\bfr).
\ee
In these equations,   $\psi$ is the Grassmann field  for a set of  transformed ``composite fermions" (CFs), whose density $\bar{\psi} \psi$ is identical to the electron density $ n_{\rm{el}}(\bfr)$.

At this stage, we have merely transformed one insoluble problem to another.  However, the transformed problem admits a sensible mean-field approximation, whereas the original problem did not.  In particular, if the Landau level is half full, so that there is one electron for each quantum of electromagnetic flux, the mean field problem describes a set of non-interacting fermions in zero magnetic field.  To go beyond mean-field theory, one must include the effects of fluctuations in the gauge field and fluctuations in the two-body potential. The central hypothesis of HLR is that, in principle,  one could obtain the correct properties of the system by  starting from the mean field solution, treating the omitted fluctuation terms via  perturbation theory.  This assumes that the interacting ground state can be reached from the mean-field solution by turning on the perturbing terms adiabatically, without encountering any phase transition.  Among the consequences of this assumption are that the ground state at $\nu = 1/2$ should be compressible, and that there should be something like a Fermi surface, with a well-defined Fermi wave vector, $k_F = 4 \pi n_{\rm{el}}$\cite{hlr,altshuler94,kim94,sternSH95}.

Experimentally, in GaAs two-dimensional electron systems, it appears that the HLR hypothesis is correct for electrons in the lowest Landau level. However, it appears that the HLR hypothesis breaks down for electrons in the second Landau level, where one observes an incompressible fractional Hall state, with an energy gap,  at half filling, in high quality samples\cite{willetteven87}.   It is widely believed that this quantized Hall state may be understood as arising from an instability of the Fermi surface to formation of Cooper pairs in the second Landau level\cite{mooreMR91,readgrn,levinapf,ssletalapf}. In still higher Landau levels, it appears that the Fermi surface is unstable with respect to the formation of charge density waves, which can lead to a large anisotropy in the measured electrical resistivity at low temperatures\cite{lillyanis99,koulakovKFS96,foglerFK97,moessnerchalker96}.

If one is interested in dynamic properties, such as the response to a time-dependent and space-dependent electric field, the first level of approximation, beyond static mean field theory, is the random phase approximation (RPA), or time-dependent Hartree approximation.  In this approximation, the composite fermions are treated as non-interacting fermions, with the bare mass $m$, driven by an effective electromagnetic field which is the sum of the applied external electromagnetic field, the Hartree potential arising from the interaction $V_2$, in the case where there are induced modulations in the self-consistent charge density, and induced electric and magnetic fields arising from modulations of the Chern Simons gauge field.  These fields may be written as
\be
\bfe =  - 4 \pi \hat{z} \times \bfj_{\rm{el}}\, , \, \,\,\,\,b = - 4 \pi n_{\rm{el}} \,,
\ee
where $\bfj_{\rm{el}}$ is the  electron current density at the point in question.

As we shall discuss further below, many properties of the system near $\nu=1/2$ are described properly by the RPA, including the response of the system to a uniform time-dependent electric field.  However, use of the unrenormalized  electron mass as assumed in the RPA, can lead to a serious error in the energy scale for various excitations.  A proper low-energy description of the composite fermion liquid requires the use of an effective mass $m^*$, which may be very different than the bare mass $m$.  In particular, one expects that the renormalized mass should be determined by the electron-electron interaction $v_2$, and should be independent of $m$, in the limit where $m \to 0$ and the cyclotron energy goes to infinity.  The renormalized mass enters directly in the low temperature specific heat, and it also is manifest in the magnitudes of the energy gaps at fractional quantized Hall states of the form $\nu= p/(2p+1)$, where $p$ is a positive or negative integer, in the limit $|p| \to \infty$ or $\nu \to 1/2$\cite{hlr,sternSH95}.

A simple modification of the RPA, which we denote RPA*,  would consist of replacing $m$ by $m^*$ in the RPA.  Although this would correctly give the energy scale for the specific heat and energy gaps in the fractional quantized Hall states, this would change the response to a time-dependent uniform electric field, which was correctly given in RPA.  Specifically, if we write
$\bf{E} = \hat{\rho} (\omega) \bfj_{\rm{el}}$, at frequency $\omega$, then it is required by Kohn's theorem that the resistivity tensor should be given by
\be
\label{rho}
\hat{\rho}(\omega) = - i m \omega  - 4 \pi \hat{\epsilon},
\ee
 where $ \hat{\epsilon}$ is the unit antisymmetric tensor, $\epsilon_{xy} = -\epsilon_{yx} = 1$.
 Using RPA*, one would find, incorrectly, that $m$ is replaced by $m^*$ in the formula for $\hat{\rho}$.

 This defect in RPA* is familiar from the theory of ordinary Fermi liquids.  In order to get the correct low-frequency response functions in the presence of a renormalized effective mass, it is necessary to include effects of the Landau interaction parameters $F_l$.  These may be defined by the energy cost to form a distortion of the Fermi surface. Specifically, a small distortion of the form
 \be
\delta k_F (\bfr, \theta) =  \sum_{l=- \infty}^{\infty}  u_l(\bfr) e^{- i l \theta}
\ee
will have an energy cost
\be
\delta E =  \frac{v_F^*k_F} {4 \pi} \int d^2 \bfr \sum_{l=- \infty}^{\infty} (1+F_l) \, \,  |u_l(\bfr)|^2 ,
\ee
where $v_F^* \equiv k_F/ m^*$.
For a Galilean invariant system, we must have
\be
F_1 = F_{-1} = (m / m^*) -1 = (v_F^* / v_F) - 1 .
\ee
As noted in SH \cite{SimonHalperin}, inclusion of these interaction parameters will also restore the correct response for the composite fermion system at $\nu=1/2$. In the presence of a non-zero current, the $l=\pm 1 $ parameters lead to an extra force on the electrons, which restores $m^*$ to $m$ in the resistivity tensor (\ref{rho}).

We remark that it is also necessary to take into account a Landau interaction parameter if one wishes to obtain the correct value for the electron  compressibility.  As in a normal Fermi liquid, we have
\be
\frac{d \mu} {d n_{\rm{el}} } = \frac{2 \pi} {m^*} (1 + F_0) ,
\ee
 where $\mu$ is the chemical potential (defined to  exclude  the contribution of the macroscopic  electrostatic potential).

 \subsection{Infrared divergences}

As was already observed in HLR, in the case of Coulomb interactions, which behave as $1/r$ for large separations $r$, an analysis of contributions to the effective mass $m^*$ arising from long-wavelength fluctuations of the Chern-Simons gauge field predicts a logarithmic divergence in $m^*$ as one approaches the Fermi surface. A similar divergence is found in the Landau interaction parameters, however, so that Galilean invariance is preserved, and the compressibility  remains finite. The decay rate for quasiparticles close to the Fermi energy is predicted to be small compared to the quasiparticle energy, in this case, so that the quasiparticle excitations remain well-defined, and the composite Fermion system may be described as a ``marginal Fermi liquid."   Similar infrared divergences are found in the Son-Dirac theory of the half-filled Landau level.

It is believed that these infrared divergences will be absent, and $m^*$ will remain finite,  if one assumes an electron-electron interaction that falls off more slowly than $1/r$, so that long-wavelength density fluctuations in the electron density are  suppressed.  Moreover, these divergences are irrelevant to the issues of PH symmetry which are the focus of the current investigation.
Consequently, we shall  assume, for the purposes of  our discussion, that we are dealing with an electron-electron interaction that falls of more slowly than $1/r$ and that  $m^*$ is finite.

We remark that for short-range electron-electron interactions, fluctuations in the gauge field lead to divergences that are stronger than logarithmic, and long-lived quasiparticles can no longer be defined at the Fermi surface.  Nevertheless, it is believed that many predictions of the HLR theory remain valid in this case\cite{kim94,altshuler94}.
 We expect that the results  of the present paper  with regard to particle hole symmetry should also apply in the case of short-range interactions, but we have not investigated this case in detail.

\subsection {Energy gaps at $\nu = p / (2p+1)$}
 \label{sec:Eg}

According to the HLR picture, if there is a finite effective mass $m^*$ at $\nu=1/2$, then for fractional quantized Hall states of the form $\nu= p /(2p+1)$, where $p$ is a positive or negative integer,  the energy gaps, in the limit $p \to \infty$, should have the asymptotic form
\be
\label{Eg}
E_g = \frac {|\Delta B|}{m^*} ,
\ee
where $\Delta B$ the deviation from the magnetic field at $\nu=1/2$ for the given electron density, i.e.,
\be
\Delta B = B - 4 \pi n_{\rm{el}} = \frac{B} {2p+1} .
\ee
Note that the allowed values of $\Delta B$ are symmetric about $\nu=1/2$,
assuming that the electron density is varied while $B$ is held fixed, since
$\Delta B(p) = - \Delta B (-p-1)$.

In the limit $m \to 0$, PH symmetry requires that the energy gap should be the same for $\Delta B $ and $- \Delta B$, assuming that  $B$ has been  held fixed.
Equation  (\ref{Eg}) will satisfy this requirement, at least to first order order in $\Delta B$. Symmetry beyond first order depends on the choice of
$m^*$  used in the formula. Although the HLR analysis  specifies that $m^*$ should be evaluated under the condition of $\nu=1/2$, there is still an ambiguity when $\Delta B \neq 0$, because one must decide whether to use the value appropriate for the given magnetic field or for the given electron density.
 These conditions are precisely  equivalent to each other only when $\Delta B = 0$.  If one employs   in
 Eq.~(\ref{Eg}) the value of  $m^*$ calculated at the given value of $B$, then the formula will exhibit PH symmetry to all orders in $\Delta B$. If one were to use the value of $m^*$ calculated at the given value of  $n_{\rm{el}} $, however, there would be violations of PH symmetry at second order in $\Delta B$.

In practice, the value of the renormalized mass cannot be calculated entirely within the HLR approach, so the value of $m^*$ to be used in the effective theory must be obtained from experiment or from some other microscopic calculation.  Thus we can say that the HLR  theory is compatible with PH symmetry in the fractional quantized Hall  energy gaps, but it can only be deduced from the theory to first order in $\Delta B$.  We remark that the same situation occurs in the Son-Dirac theory.  Precise PH symmetry in that case depends on a separate assumption that the renormalized value of the Dirac velocity should be  determined by the magnetic  field and not by the electron density.

\section{DC transport at $\nu = 1/2$  \label{sec:dc}}

PH symmetry, in the limit $m \to 0$, implies that the  Hall conductivity  in response to a spatially uniform electric field should be precisely given by
\be\label{sig}
\sigma_{xy} = - \sigma_{yx} = \frac {1} {4 \pi},
\ee
regardless of the applied frequency.  This should be true even in the presence of impurities, provided that the disorder potential $V_{\rm{imp}}$ is PH symmetric in a  statistical sense.  This means that if one chooses the  uniform background potential such that the average  $\langle V_{\rm{imp}} \rangle = 0$, then all odd moments of the disorder potential must vanish.

In the absence of impurities, we may use the result (\ref{rho}) for a Galilean invariant system to calculate the conductivity tensor
\be\label{sig2}
\hat{\sigma}(\omega) = \hat{\rho}^{-1}(\omega) = \frac {-  i m \omega +4 \pi \hat{\epsilon}}
{ 16 \pi^2- m^2 \omega^2 }  .
\ee
If $m=0$, this gives $\hat{\sigma}(\omega) =  - \hat{\epsilon} / 4 \pi$, which satisfies the condition for PH symmetry.  As we have seen, the HLR theory  will satisfy Galilean invariance if the
$F_{\pm 1}$ interaction parameter is taken into account.  However, if one were to use the renormalized mass without the $F_{\pm 1}$ interaction, one would  find that $m$ is replaced by $m^*$ in Eq. (\ref{sig2}), so that particle hole symmetry would not be satisfied for $\omega \neq 0$.

Of greater interest is the dc Hall conductivity in the presence of impurities. For many years, beginning with the work of Kivelson et al. in 1997\cite{klkgphhlr}, it has been widely believed that the HLR approach must give
a result for the $dc$ Hall conductivity that is inconsistent with PH symmetry, at least at the level of RPA and perhaps beyond, if the mean free path for composite fermions is finite. The reasoning goes as follows.  Within the HLR approach, the electron resistivity  tensor is related to the resistivity tensor of the composite fermions by
\be
\hat{\rho} = \hat {\rho}^{\rm{\, cf}} +   \hat {\rho}^{\cs} ,
\ee
where  $\hat {\rho}^{\cs}$ is the Chern-Simons resistivity tensor, given by
\be
 \hat {\rho}^{\cs} =  - 4 \pi \hat{\epsilon} .
 \ee
One finds that in order to obtain the PH symmetric result for $\sigma_{xy}$, if $\rho_{xx} \neq 0$, it is necessary that $\sigma^{\rm{\, cf}}_{xy }=  - 1 / 4 \pi.$  However, it was argued that $\sigma^{\rm{\, cf}}_{xy}$ is necessarily equal to zero at $\nu=1/2$.  This is because, in the absence of impurities, the composite fermions see an average effective magnetic field equal to zero, which is effectively invariant under time reversal. The presence of impurities leads to non-uniformities in the electron density, which lead to local fluctuations in the effective magnetic field $b(\bfr)$. These fluctuations,  in turn, will be the dominant source of scattering of composite fermions, under conditions where the correlation length for the impurity potential is large compared to the Fermi wave length.  If the impurity potential is statistically PH symmetric, then there will be equal probability to have a positive or negative value of $b$  at any point, so that the resulting perturbation to the composite fermions should again be invariant under time reversal in a statistical sense.

The fallacy we find here in this reasoning is that fluctuations in $b$ are correlated with fluctuations in the electrostatic potential, which though their effects are  weak compared to the effects of $b$, are sufficient to break the statistical time-reversal symmetry produced by the $b$ fluctuations alone. We shall see below that when these correlated fluctuations are taken into account we recover precisely the result
$\sigma^{\rm{\, cf}}_{xy}= - 1 / 4 \pi$ required by PH symmetry.

In the subsections below, we show how disorder leads to the desired result for
$\sigma^{\rm{\, cf}}_{xy }$.  As there are some subtleties involved in these calculations, we present here two  different derivations, which bring different insights to the problem and which may be  applicable in somewhat different regimes.  The first derivation employs a semi-classical analysis and  
uses the Kubo formula, which expresses the conductivity in terms of equilibrium correlation functions. The second derivation employs the Born Approximation and the Boltzmann Equation , and calculates the conductivity by analyzing the effect of the electric field on the particles' dynamics. We also discuss consequences for thermoelectric transport at $\nu=1/2$.  

Our calculations are restricted to the case where the Fourier components of the disorder potential have wave vectors small compared to $k_F$. Neither the HLR nor the Dirac theories, in their simplest forms, can describe quantitatively the effect of potential fluctuations with wave vectors comparable to or larger than $k_F$. In either theory,  the coupling between a short-wavelength potential fluctuation and the operators that scatter a composite fermion from one point to another on the Fermi surface will be affected by vertex corrections,
whose value is determined by microscopic considerations and cannot be calculated within the low-energy theory itself.

It should be emphasized that while the effects discussed below may be important as a matter of principle, they are all sub-leading corrections to the transport in the presence of impurities.  For small impurity concentrations, the CF Hall conductance $\sigma^{\rm{\, cf}}_{xy }= - 1 / 4 \pi$ is  small compared to the diagonal CF conductance, $\sigma^{\rm{\, cf}}_{xx }$, which is proportional  to $k_F l_{\cf}$, where $l_\cf$ is the transport mean free path for composite fermions. 
 Conversely, if one were to set $\sigma^{\rm{\, cf}}_{xy }= 0$, this would lead to a deviation of the electronic $\sigma_{xy}$ from the PH-symmetric value by an amount proportional to  $\sigma_{xx}^2 \propto
1/(k_F l_{\cf})^2$, which is small compared to $\sigma_{xx}$ as well as  to $\sigma_{xy}$, in the limit  of large $k_F l_{\cf}$.

\subsection{Disorder potential and fluctuations of the magnetic field} \label{sec:Dpm}

In general, density fluctuations produced by an external electrostatic potential such as $V_{\rm{imp}}$ will tend to screen the external potential and give rise to a combined self-consistent potential,  which we denote $V(\bfr)$.  Within a mean-field approximation, for long-wavelength potential fluctuations, the induced density fluctuation should be related to $V$ by
\be
\delta n^{\cf} (\bfr) = - \chi V(\bfr) ,
\ee
where $\chi$ = $m / 2 \pi$ is the compressibility of noninteracting fermions. We assume here that the potential  $V_{\rm{imp}}$ contains only Fourier components with wave vectors $q$ that are small compared to $k_F$, which is appropriate for a remotely doped system,  where the impurities are set back from the 2DES by a distance large compared to the Fermi wavelength.

Beyond the mean field approximation, we should replace  $m$ by $m^*$, and we should redefine the potential $V$ to include effects of the $F_0$ Landau parameter.
The effective magnetic field $\delta b  = \langle b(\bfr) \rangle + B$ produced by a fluctuation in the redefined $V$ is then given by
\be
\label{bV}
\delta b(\bfr) =  2 m^* V(\bfr)\, .
\ee
Equivalently, we may describe this in terms of the induced vector potential $\delta \bfa$, which may be written in Fourier space as

\be
\delta \bfa (\bfq) = -  2  m^* V (\bfq)   \frac {i \hat{z}\times \bfq}{q^2}
\ee
Since the gauge fluctuation will couple to the momentum of a composite fermion with a term $- \delta \bfa \cdot \bfp_j / m^*$, we find that the total effect of the impurity potential is a term in the Hamiltonian whose matrix element between an initial state of momentum $\bfk'$ and a final state $\bfk $ is given by
\be
\label{Hkk}
U_{\bfk \bfk'}  = V(\bfq) \left[ 1 + \frac { 2 i (\bfk \times \bfk') \cdot \hat{z} }{q^2} \right] ,
\ee
where $\bfq = \bfk - \bfk'$.

\subsection {Semiclassical analysis using the Kubo formula}

In this subsection, we employ a semiclassical analysis of the dynamics of CFs of mass $m^*$ in the presence of the (screened) impurity potential $V({\bf r})$. We restore factors of $e$ and $\hbar$, and we consider a more general situation, where $\nu = 1 / (2n)$, where $n$ is an integer, not necessarily equal to 1. Then Eq (\ref{bV}) for the effective magnetic field $\delta b $ should be replaced by
\be
\delta b = V(\bfr) \frac {2 n m^*} {\hbar e}  .
\ee

The semiclassical equations of motion are then
\begin{eqnarray}
\label{eq:pdot}
\dot{{\bf p}} & = & -{\bf \nabla} V +\frac{2 n V({\bf r})}{\hbar} {\bf p}\times \hat{\bf z} \\
\dot{{\bf r}} & = & {\bf p}/\m \,.
\end{eqnarray}
(We assume, here, and in the formulas below,
that the product of the electron charge and the $z$-component of the external magnetic field is positive. Results for the  opposite case may be obtained by interchanging  indices for the $x$ and $y$ axes.)

We shall consider $V(\bf r)$ to be a random function, symmetrically distributed around $V=0$. Its  correlation length $\xi$ is assumed large compared to  $\hbar/p_F$ with $p_F = \sqrt{2\m \varepsilon}$ the Fermi momentum and $\varepsilon$ the Fermi energy,  as required for validity of the semiclassical approximation.
Note that the Lorentz force (of order $Vp_F/\hbar$) is then large compared to the force exerted by gradient of the potential (of order $V/\xi$) by a factor $\xi p_F/\hbar$.  The validity of the semiclassical analysis also requires that the typical scattering angle from this Lorentz force, $\Delta \theta \sim V m^* \xi /(\hbar p_F)$, is large compared to the diffraction angle $\sim \hbar / (\xi p_F)$,  i.e.  $V \gg \hbar^2 / (m^* \xi^2)$.

It is convenient to separate into radial and angular co-ordinates, by writing
\begin{equation}
p(t) \equiv p_x(t)+ i p_y(t) = |p(t)|e^{i\theta(t)} \,.
\end{equation}
For a particle of energy $\varepsilon$
\begin{equation}
\label{eq:p}
|p(t)| = \sqrt{2\m\{\varepsilon - V[{\bf r}(t)]\}}
\end{equation}
while the angle $\theta$ must be found by integrating
\begin{equation}
\label{eq:thetadot}
\dot{\theta}(t) = \frac{1}{|p(t)|} \left(\sin\theta \frac{\partial V}{\partial x} - \cos\theta \frac{\partial V}{\partial y}\right) - \frac{2n V}{\hbar}
\end{equation}
along the  trajectory ${\bf r}(t)$ of the particle.

We shall use the classical form of the Kubo formulas for the conductivity in terms of velocity-velocity correlation functions.
To this end, we  construct the correlator
\begin{eqnarray}
\label{eq:correlator}
K(t-t_0) & \equiv & \frac{1}{{\m}^2}\left\langle p(t)p^*(t_0) \right\rangle
\end{eqnarray}
with the average taken over the distribution of particles in phase space.  To represent the degenerate Fermi gas we shall consider the microcanonical
distribution at the Fermi energy $\varepsilon$. The conductivities are then
\begin{equation}
\label{eq:kubo}
\sigma^\cf_{xx} - i \sigma^\cf_{xy} = \frac{\m}{2\hbar}\left(\frac{e^2}{h}\right) \int_0^\infty K(t) \, dt
\end{equation}
where the prefactor involves the compressibility.
For fixed Fermi energy $\varepsilon$, large compared to $V$, we use (\ref{eq:p}) expanded to first order in $V/\varepsilon$, to write
\begin{eqnarray}
K(t-t_0)
 &  \simeq & \frac{2}{\m}\left\langle \left[\varepsilon -  \frac{V({\bf r}_t)+ V({\bf r}_{t_0})}{2}\right]
e^{i\int_{t_0}^t\dot\theta(t')dt'}
 \right\rangle   \nonumber \\
 \end{eqnarray}
 and then use (\ref{eq:thetadot}) to  replace
$V({\bf r})\simeq -(\hbar/   2n  )\dot\theta$ for $\xi p_F/\hbar\gg 1$ at both $t$ and $t_0$,
leading to
\begin{eqnarray}
\label{eq:final}
 K(t-t_0)     &  \simeq &   \frac{2}{\m}  \left[\varepsilon - \frac{i\hbar}{2n }\frac{d}{dt}\right]\left\langle
 e^{i\int_{t_0}^t\dot\theta(t')dt'}
\right\rangle \,.
\end{eqnarray}

The correlator
\begin{equation}
\label{eq:average}
\left\langle
e^{i\int_{t_0}^t\dot\theta(t')dt'}
\right\rangle
\end{equation}
depends on how the particles move in real space.
Assuming that the composite mean free path $l_{\cf}$ is large compared to the correlations length $\xi$ for fluctuations in the potential $V$, we may expect that
each particle will explore phase
space with the probability of the microcanonical
distribution, $\rho({\bf p},{\bf r}) \propto \delta[\varepsilon - |{\bf p}|^2/2\m-V({\bf r})]$.  (The assumption
$l_{\cf} \gg \xi$ is clearly valid in the limit where the magnitude of the potential fluctuations is small while $\xi$ is held fixed.)
Integrating the microcanonical distribution over 2D momentum  leads to a uniform real-space density distribution [since $\varepsilon >V({\bf r})$]. Thus,
each particle moves in such a way that its time-varying potential
${V}[{\bf r}(t)]$ has the same probability distribution as $V({\bf
  r})$.
For example, from Eqn (\ref{eq:thetadot}),  $\dot\theta$ vanishes under time-averaging.
More specifically,  since the distribution of $V$ is invariant under $V\to-V$, so too is that of ${\dot\theta}$ under $\dot\theta\to -\dot\theta$, such that (\ref{eq:average}) is real.
Hence, from (\ref{eq:final})
\begin{eqnarray}
{\mathcal Im}\left[K(t) \right] & \simeq & -\frac{\hbar}{n \m}\frac{d}{dt}\left\langle
e^{i\int_{0}^t\dot\theta(t')dt'}
\right\rangle \,.
\label{eq:imk}
\end{eqnarray}
Inserting this in Eqn (\ref{eq:kubo}), and noting that  the correlator (\ref{eq:average}) will vanish at $t-t_0\to \infty$
for any  disordered potential, we  obtain the result
\begin{eqnarray}
\label{scresult}
\sigma^\cf_{xy}
& = &- \frac{1}{n }\left(\frac{e^2}{ 4 \pi   \hbar }\right) \,.
\end{eqnarray}

For the case $\nu=1/2$, where $n=1$,  we recover our desired result $\sigma^\cf_{xy} = - 1/(4 \pi)$, in units where $e = \hbar =1$. More generally, the result (\ref{scresult}) implies that the electron Hall conductivity at $\nu = 1/ (2n)$ is precisely given by $\sigma_{xy} = (e^2)/ (4 \pi \hbar  n)$, even in the presence of impurities. Thus there seems to be  a kind of emergent PH symmetry at fractions such as $\nu=1/4$ and $\nu=1/6$.

\subsection {Calculation using the Born Approximation and Boltzmann Equation}

It seems reasonable that we are  justified in using a semiclassical approximation for our problem, because we are necessarily focused on potential fluctuations on a length scale $\xi$ that is large compared to  $k_F^{-1}$.  However, the requirement also that the classical scattering angle exceeds the diffraction angle, [{\it i.e.}, the condition  $V \gg \hbar^2/ (m^* \xi^2)$ discussed above], leads to some subtleties in the applicability of the classical results for weak potentials\cite{dyakonovkhaetskii}. 
It can be shown that the transport scattering cross section, (i.e., the  integrated cross section weighted by the square of the momentum transfer) is correctly given by the semiclassical approximation in this case, and it agrees with a quantum mechanical calculation based on the Born approximation.  However, the total scattering cross section, as well as the differential cross section at any particular angle,  is generally not given correctly by a semiclassical analysis.  Therefore, it seems useful to check that our semiclassical calculation of the off-diagonal part of the CF conductivity tensor can be duplicated in a more quantum mechanical calculation.

Here we follow closely the analysis used by Nozi\`eres and Lewiner (NL)\cite{NL73}  for the anomalous Hall effect due to spin-orbit interactions in a spin-polarized semiconductor.  In their analysis, NL employed a Boltzmann equation to study the evolution of the electron system in a uniform applied electric field, paying careful attention to the effects of spin orbit coupling on the collision integral in the presence of the field.

In our case, we wish to study carefully the scattering of a composite fermion by an  impurity described by an effective Hamiltonian  of the form (\ref{Hkk}).  In order to use the NL analysis directly,  we must impose the condition that the scattering matrix element $U_{\bfk \bfk'}$ due to a single impurity is zero in the limit $\bfk \to \bfk'$.  This means that the associated  potential $V(\bfq)$ should vanish for $\bfq \to 0$ faster than $q$.  In real space, this means that the space integral of the potential $V(\bfr)$ should vanish, as well as its first spatial moments.  If individual impurities do not satisfy these conditions, the NL analysis may still be used if impurities can be grouped into small clusters that satisfy the conditions.  In any case, the purpose of this subsection is to provide a check of the validity of the above-described semiclassical approximation as a matter of principle, rather than to check the validity in a realistic situation.

It is instructive to describe our calculation in two parts. In the first part we consider the scattering of a single composite fermion from momentum $\bfk$ to momentum $\bfk'$ by the potential (\ref{Hkk}) in the absence of an electric field. We show - following NL - that this scattering involves a ``side-jump" $\delta \bfr_\bfq =  - \frac {(\hat{z} \times \bfq) }{ (2 k_F^2)}$, i.e., a motion of the electron in the direction perpendicular to the momentum transferred from the disordered potential to the composite fermion. When averaged over all scattering processes  from a momentum $\bf k$ each scattering event involves a side-jump,
which 
results in a net motion perpendicular to the direction of $\bfk$. In the presence of an electric field $E_x$, the net flux of electrons that experience scattering by the potential is proportional to $e E_x\tau$, where $\tau \equiv l_{\cf} m^*/k_F$ is the transport scattering time.  As they scatter from impurities,  the extra electrons acquire a velocity in the $y$-direction given by  $\Delta/\tau$ where $\Delta$ is the 
cumulative side jump during the time $\tau$ in which their direction of motion is randomized. Since $\Delta$ is of order $k_F^{-1}$, this results in a
current in the $y$-direction of the order of $\frac{e^2}{h}E_x$, which  gives rise to a non-zero contribution to $\sigma_{xy}^{\cf}$ that is independent of the mean free path.

In the second part we consider the effect of an applied electric field on the scattering. In the presence of that field the change in position associated with the side-jump implies that the scattering of the composite fermion involves also a change in its  kinetic energy. As explained below, that change results in another contribution to the Hall current, equal in magnitude and sign to the first contribution.  Throughout this subsection, we assume $n=1$, and return to  units where $e=\hbar=1$.

\subsubsection{Scattering rate of a single composite fermion}

For the first part, suppose that a composite fermion,  described by a Gaussian wave packet, centered at a momentum $\bfk_0$ on the Fermi surface, is incident on the impurity.   As discussed in Appendix B of NL, we may write the wave function of the CF as
\be
\psi(\bfr, t) = \sum_{\bfk} C_{\bfk} (t) e^{i \bfk \cdot \bfr}
\ee
\be
C_{\bfk} = C^0_{\bfk} + C^1_{\bfk} + C^2_{\bfk} ,
\ee
where $C^0_{\bfk}$ describes the incident wave:
\be
C^0_{\bfk} = N e^{-i \varepsilon_{\bfk} t } e^{- (\bfk - \bfk_0)^2 /2 \Delta^2  }
\ee
where $\varepsilon_{\bfk}$  is the energy of a fermion of wave vector $\bfk$, and $N$ is a normalization constant, and $C^1$ and $C^2$ are of order $U$ and $U^2$ respectively. (Note that the incident wave packet is centered at the origin at time $t=0$.) In the limit of large positive times one finds that
\be
C^1_{\bfk}= - 2 \pi i \sum_{\bfk '} U_{\bfk \bfk'} \delta (\varepsilon_{\bfk}-\varepsilon_{\bfk'})
C^0_{\bfk '}
\ee
\begin{eqnarray}
C^2_{\bfk}  &=&   - 4 \pi^2 \sum_ {\bfk' \bfk ''}  U_{ \bfk \bfk''}  U_{ \bfk''  \bfk'}
\delta (\varepsilon_{\bfk}-\varepsilon_{\bfk''}) \times \\ \nonumber
& &  \times \delta (\varepsilon_{\bfk''}-\varepsilon_{\bfk'}) C^0_{\bfk '} .
\end{eqnarray}

As noted in NL, the average position of the particle at time $t$ can be written as
\be
\frac{i}{2} \sum_\bfk \left[C^*_\bfk \frac {\partial C_\bfk} {\partial \bfk}
-C_\bfk \frac {\partial C^*_\bfk} {\partial \bfk}   \right]
= \sum_\bfk |C_\bfk |^2 \bfr_\bfk ,
\ee
where
\be
\bfr_\bfk =  - \frac{\partial}{\partial \bfk} {\rm{Arg}}\,  C_\bfk .
\ee
There are two contributions to the shift of the average position. The first is seen when we consider a momentum $\bfk$ in the scattered wave, with $|\bfk - \bfk_0| \gg \Delta$,  so that  $C^0_\bfk = 0$. Then, to lowest order, $C_\bfk$ may be replaced by $C^1_\bfk$, and the phase is equal to the phase of $C^1$. Using (\ref{Hkk}) for $U$, we find that $C^1$ has an extra argument, beyond the contribution from $e^{- i \varepsilon_\bfk t}$, arising from the complex value of $U_{\bfk \bfk '}$. This extra argument has the form
${\rm{Arg}} \, C^1_\bfk  \sim   - q^2 / [ 2 \hat{z} \cdot (\bfk \times \bfk_0)]      $,
and it leads to an extra displacement of the center of the scattered wave packet by an amount
\be
\delta \bfr^{(1)}_\bfk = -  \frac{\hat{z} \times \bfk }{2 k_F ^2} .
\ee
  The second contribution to the average displacement comes from weight that has been asymmetrically  removed from the incident part of the wave packet, where $\bfk$ is close to $\bfk_0$.  Here there is an interference between $C^0$ and $C^2$. If one assumes that $V(\bfq)$ is vanishing for $\bfq = 0$, then one finds that the contribution from this term is given by
  \be
  \delta \bfr_0 =  \sum_\bfk | C^1_\bfk |^2 (\hat{z} \times \bfk) / (2 k_F ^2) .
  \ee
Summing the two contributions we find that the net displacement (``side jump") associated with a particle that scatters from a direction $\bfk_0$  into direction $\bfk = \bfk_0 + \bfq$ depends on the transferred momentum, and  is  given by 
\be
\label{sidejump}
\delta \bfr_\bfq =  - \frac {(\hat{z} \times \bfq) }{ (2 k_F^2)}.
\ee
This side jump contributes directly to the total current through a net charge displacement per unit time
\be
\label{dJ}
\delta\mathbf{J}=\sum_{\bfk,\bfk'}f(\bfk)W_{\bfk,\bfk'}\delta\bfr_{\bfk'-\bfk},
\ee 
where $f(\bfk)$ is the occupation  probability for  a state of momentum $\bfk$, and $W_{\bfk, \bfkÕ}$ is the transition probability [see Eq.~(\ref{scatteringrate}) below]. We can express the side-jump contribution $\bf \delta J$ in terms of  the current in the absence of that contribution, ${\bf J}_0=\sum_{\bf k}f({\bf k})k/m^*$. Using Eq.~\eqref{sidejump} for the displacement, and noticing that the transport scattering rate is given by 
\begin{equation}
\frac{1}{\tau}\equiv\sum_{\bfk'}W_{\bfk,\bfk'}(1-{\hat k}\cdot{\hat k'})
\label{oneovertau}
\end{equation}
we can simplify  Eq.~(\ref{dJ}) to 
\be
\delta\mathbf{J}=-\frac{m^*}{2\tau k_F^2}\mathbf{J}_0\times\hat{z}.
\ee
Since,  to leading order, $\mathbf{J}_0=\frac{n_e\tau}{m^*}\mathbf{E}$, the $\delta\mathbf{J}$ term leads to a contribution to $\sigma_{xy}^{\cf}$ of the form
\be
\sigma^{\rm{sj}}_{xy} = -  \frac {1} {8 \pi} .
\label{sidejumpcontribution}
\ee

\subsubsection{Scattering rate in a composite fermion liquid in the presence of an electric field}

Eq. (\ref{sidejumpcontribution})  is half of the amount we need for PH symmetry.
The second half is a consequence of  having a liquid of composite fermions, in which an applied electric field affects the occupation of momentum states. While the scattering rate from momentum $\bfk$ to momentum $\bfk'$ is symmetric with respect to the sign of $(\bfk\times\bfk')\cdot {\hat z}$ for a single composite fermion in the absence of an electric field, the situation is more complicated in the presence of both a liquid of composite fermions and an electric field.
In that case, due to the electric field the side-jump is associated with a change of the composite fermion's kinetic energy by an amount $e{\bf E}\cdot{\delta {\bf r}_{\bf q}}$. 
The effect of this change of energy on the transport is best understood by means of the Boltzmann equation. For $dc$ transport in the presence of impurities the equation reads
\begin{equation}
\bfF \cdot \nabla_\bfk f = -\sum_{\bfk'}W_{\bfk,\bfk'}(f(\bfk)-f(\bfk')) ,
\label{Boltz}
\end{equation}
where
\begin{equation}
W_{\bfk,\bfk'}=2\pi |V_{\bfk,\bfk'}|^2\delta(\epsilon_\bfk+\bfF\cdot\delta {\bfr}_{\bfq}-\epsilon_{\bfk'})
\label{scatteringrate}
\end{equation}
Here $f_0$ is the Fermi-Dirac distribution, $\bfF=e{\bf E}$ is the force acting on the composite fermions, $\epsilon$ is the energy, and $V$ is the disordered potential. The $\delta$-function expresses the change of the kinetic energy incurred by the scattered electron, a change which is our main focus here.

As customary, linear response to $\bfF$ is analyzed by setting $f$ to be $f_0$ on the left-hand side of (\ref{Boltz}) and by 
writing $f(\bfk)=f_0
+f_1=f_0+\frac{\partial f_0}{\partial\epsilon}{\bf u}\cdot{\bfv}_{\bfk}$ on the right-hand side. The transfer of energy affects the expansion of the distribution functions on the right hand side. Specifically we have, 
\begin{eqnarray}
&\sum_{\bfk'}W_{\bfk,\bfk'}(f(\bfk)-f(\bfk'))=\nonumber \\ 
&-\frac{\partial f_0}{\partial\epsilon}{\bf u}\cdot\sum_{\bfk'} \left[   W_{\bfk,\bfk'}(\bfv(\bfk)-\bfv(\bfk'))+\frac{\partial f_0}{\partial\epsilon}W_{\bfk,\bfk'}\bfF\cdot\delta r_q \right ]. \nonumber
\label{Boltzexpansion}
\end{eqnarray}

We now make use of the definition of the transport scattering rate (\ref{oneovertau})
to write the Boltzmann equation as 
\begin{equation}
\bfF\cdot \left(\bfv_\bfk-\frac{{\hat z}\times\bfk}{2k_F^2\tau} \right)  \frac{\partial f_0}{\partial\epsilon}=\frac{{\bf u}\cdot\bfv_\bfk}{\tau}\frac{\partial f_0}{\partial\epsilon}  ,
\end{equation}
which amounts to 
\begin{equation}
f_1({\bf k})=\tau\bfF\cdot \left(\bfv_\bfk-\frac{{\hat z}\times\bfk}{2k_F^2\tau}\right)\frac{\partial f_0}{\partial\epsilon} .
\label{fone}
\end{equation} 
As this expression shows, in the limit of a small scattering rate $1/\tau$ the shift of the Fermi sea that results from the application of the electric field is primarily parallel to the electric field, but includes also a small term perpendicular to the field. This term contributes to the Hall conductivity. 

The current is ${\bf J}=\int d\bfk f_1(\bfk)\bfv_\bfk$, with $\int d\bfk=\frac{m^*}{(2\pi)^2}\int d\epsilon \, d\theta$. The angular integral gives $\pi$ for both components of the current (each component from a different term), leading to $\sigma_{xx}^{\cf}=\frac{k_F v_F \tau    }{4\pi}$, and $\delta \sigma_{xy}^{\cf}=-\frac{1}{8\pi}$. This contribution to the Hall conductivity adds to the side-jump contribution calculated in the previous subsection, with the sum of the two being $-\frac{1}{4\pi}$.

\subsection {Thermopower and thermal transport}

\subsubsection{General considerations}

In this subsection, we again restore $\hbar$ and  the electron charge $e$.  The formulas are correct for either sign of $e$, provided
 that the product of the electron charge and the $z$-component of the external magnetic field is positive. For $eB<0$, the  $x$ and   $y$ axes should be interchanged.

The thermoelectric and thermal responses for the CFs can be obtained from standard results for non-interacting fermions, based on interpreting the CF conductivity in terms of an energy-dependent conductivity $\Sigma^\cf_{\mu\nu}(\varepsilon)$ through
\begin{equation}
\label{eq:cond}
\sigma^\cf_{\mu\nu} = \int \Sigma^\cf_{\mu\nu}(\varepsilon) \left(-\frac{\partial f}{\partial \varepsilon}\right)
d \varepsilon ,
\end{equation}
with $f$ the Fermi distribution. We explore the consequences, making use only of the fact that  $\sigma^\cf_{xy}=-(e^2/4 n \pi \hbar)$, independent of the Fermi energy, and hence that $d\Sigma^\cf_{xy}/d\varepsilon = 0$. Here we focus on the $\nu=1/2$ state with $n=1$.

Although observations of thermal effects require that the temperature should not be too small, the calculations here also assume that the temperature  should not be too high. In particular, {\em we assume that the temperature is sufficiently low that the mean free path for inelastic scattering  of composite fermions is large compared to the mean free path for elastic scattering by impurities.}  This restriction becomes more severe as the sample becomes more ideal.

\subsubsection{Thermopower}

The heat current ${\bf j}^Q = {\bf j}^E - \mu {\bf j}^N$ induced by a field ${\bf {\mathcal E}}^\cf$ applied to the CFs
is described by a response function,
$j^Q_\mu = L^\cf_{\mu\nu}{{\mathcal E}^\cf_\nu}$, assuming that the temperature is a constant.  For a non-interacting Fermi gas, at low temperatures, expanding around the Fermi level leads to the general result
\begin{equation}
L^\cf_{\mu\nu} = \frac{\pi^2 k_B^2 T^2}{3e} \frac{d\Sigma^\cf_{\mu\nu}}{dE} \,.
\end{equation}
Since the Hall conductivity of the CFs is fixed to $\sigma^\cf_{xy} = -e^2/2h$, requiring $d\Sigma^\cf_{xy}/dE = 0$, then
\begin{equation}
L^\cf_{\mu\nu} = L^\cf_{xx} \delta_{\mu\nu} .
\end{equation}
This (diagonal) result is of the form required by PH symmetry, as discussed in \cite{PSV}, so that
 $\sigma^\cf_{\mu\nu}$ and $L^\cf_{\mu\nu}$ are each characterized by a
single non-universal quantity, $\sigma^\cf_{xx}=\sigma^\cf_{yy}$ and
$L^\cf_{xx}=L^\cf_{yy}$.

To construct the thermoelectric response tensor for
the {\it electrons} (not the CFs), one must take account of the fact that the electric field that couples to the electrons is
\begin{equation}
\bf{E} = {\bf {\mathcal E}}^\cf + \hat{ \rho}^{\cs} {\bf j}
\end{equation}
where  ${\bf j}$ is the current of either electrons or CFs and
\begin{equation}
\hat{\rho} ^{\cs} \equiv  - 4 \pi \frac{\hbar}{e^2} \hat{\epsilon} .
\end{equation}
The response tensors for the electrons are readily found to be
\begin{eqnarray}
\hat {\sigma} & = &  \hat{\sigma}^\cf (1+  \hat{\rho}^{\cs} \hat{\sigma}^\cf  )^{-1}
\\
\hat{L}  & = &  \hat{L}^\cf (1+  \hat{\rho}^{\cs} \hat{\sigma}^\cf  )^{-1}
  \,.
\end{eqnarray}
With  our specific forms of $\hat{\sigma}^\cf$ and $\hat{L}^\cf$,  these become
\begin{eqnarray}
\label{eq:sigmae}
\hat{\sigma} & = &
\frac {e^2}{4 \pi \hbar} \left[ \hat{\epsilon} + \frac{e^2} {4 \pi \hbar \sigma^\cf_{xx}}\right] ,
\end{eqnarray}
\be
\hat{L} = \hat{\epsilon} \frac {e^2 L^\cf_{xx}} {4 \pi \hbar \sigma^\cf_{xx}}  .
\ee

In a thermopower experiment, one measures a voltage gradient induced when there is a heat current, but no electric current, flowing through the sample.  Making use of an Onsager relation\cite{cooperCHR97},  as well as the relations between CF and electron coefficients, one finds  
\be
\label{tptensor}
S_{\mu\nu}  =    \frac{1}{T}\left[ \hat{L} \, ( \hat{\sigma}^\cf)^{-1}\right]_{\mu\nu}  =    \frac{1}{T}L^\cf_{xx}\left[( \hat{  \sigma}^\cf)^{-1}\right]_{\mu\nu}\,.
\ee
We see that the thermopower tensor 
has non-zero off-diagonal elements, since $\sigma^\cf_{\mu\nu}$ is not diagonal. This contrasts with predictions based on a naive application of the HLR theory, pointed out by \cite{PSV}, in which the off-diagonal thermopower vanishes. It recovers the central result of their PH symmetric theory.

\subsubsection{Thermal Transport}

In a thermal transport experiment, one seeks to measure the heat current  ${\bf j}^Q$ induced by a temperature gradient $\nabla T$, under conditions where the electrical current is zero.    As shown in Ref.~\cite{cfltislrev},
the diagonal thermal conductivity $K_{xx}$ at $\nu=1/2$ should be related by the Wiedemann-Franz law to the conductivity of the composite fermions, that is 
\begin{eqnarray}
K_{xx}
 & = & \sigma^\cf_{xx} \frac{\pi^2k_B^2T}{3e^2} .
\end{eqnarray}  
This result is obtained in both the HLR theory and the  Dirac theory.   Note that the thermal conductivity will become large as the mean free path becomes large, while the diagonal electrical conductivity 
$\sigma_{xx}$ approaches zero in this limit.

It was also suggested in Ref.~\cite{WS16} that for a system confined to  the lowest Landau level, with a particle-hole symmetric distribution of impurities, the off-diagonal thermal conductivity should be given precisely by 
\begin{equation}
K_{xy} = \frac{1}{2}\frac{\pi^2k_B^2T}{6 \pi \hbar } =   \sigma_{xy}\frac{\pi^2k_B^2T}{3e^2}  \,.   
\end{equation}
However, in an actual experiment in a strong magnetic field, one expects that  thermal gradients and currents will be quite inhomogeneous, and a major part of the thermal Hall current will be associated with chiral heat flow near the sample boundaries, where particle-hole symmetry is strongly broken\cite{cooperCHR97}.   Moreover, the transverse heat flow will be small compared to the longitudinal heat current, if the disorder scattering is weak.  A proper analysis of the transverse heat flow is, therefore,  a non-trivial problem, which we shall not address here.

\section{Commensurability Oscillations}
\label{sec:commens}


An important property investigated in HLR, which turns out to be sensitive to PH symmetry, was the wavevector dependent longitudinal conductivity, $\sigma_{xx}( q)$, for a wave vector $\bfq$ in the $x$-direction, in the limit of frequency  $\omega \to 0$. Precisely at $\nu=1/2$, In the absence of impurities, it was found, using the RPA that
\be
\label{sigxxq}
\sigma_{xx}( q) = \frac {q}{8 \pi k_F},
\ee
independent of the renormalized mass or the bare mass.   Subsequent analyses supported the idea that this result should be correct to all orders in perturbation theory, even in the case of short range electron-electron interactions or of $1/r$ interactions, where the effective mass is found to diverge at the Fermi energy\cite{kim94}.  In the presence of disorder, it was predicted that Eq  (\ref{sigxxq}) should hold for $q l_{\cf} \gg 1$,
where $l_{\cf} $ is the transport mean free path for the composite fermions.  For  $q l_{\cf} \ll 1$, the electrical conductivity approaches a constant, given by
\be
\sigma_{xx}(q=0) \approx \frac{1}{4 \pi k_F l_{\cf}}.
\ee
(This equation may be taken as a definition of $l_{\cf}$).

The non-trivial $q$-dependence of $\sigma_{xx}$ results from an  inverse $q$-dependence of the transverse conductivity for composite fermions, which is non-local, because at $\nu=1/2$, the composite fermions can travel in straight lines for distances of the order of $l_{\cf}$, which can be very large compared to the inter-particle distance $k_F^{-1}$. For filling factors that differ slightly from $\nu=1/2$, the composite fermions will no longer travel in straight lines, but rather should follow cyclotron orbits with an effective cyclotron radius given by
\be
R_C^* = \frac {k_F}{ |\Delta B|} .
\ee
One would expect, therefore, that the conductivity should become independent of $q$ for wavelengths large compared to $R_C^*$, or $q R_C^* \ll 1. $ Analysis at the RPA level, using  a semiclassical description of the composite fermion trajectories, found that the value of the conductivity in this regime is essentially the same as the $q=0$ conductivity at $\nu=1/2$.  By contrast, in the regime $q R_C^* \geq 1$, if $l_{\cf} \geq R_C^*$, one finds that the longitudinal conductivity depends on  $q$ and  $|\Delta B|$, and is a non-monotonic function of these variables. If either $q$ or $\Delta B$ is varied, one finds a series of maxima and minima, with the maxima occurring roughly at points which satisfy Eq.  (\ref{qn1}), or equivalently
\be
\label{qn2}
q R_C^* \approx  z_n .
\ee
Since $z_n \approx \pi (n+\frac{1}{4})$, with a high degree of accuracy, it is natural to describe the oscillatory dependence as a commensurability phenomenon, with maxima in $\sigma_{xx}$ where the diameter of the cyclotron orbit is approximately $(n+1/4)$ times the wavelength $2 \pi / q$.  The calculated peaks and valleys are generally broad if $q l_{\cf}$ is of order unity, but the peaks are predicted to become sharp, and the positions of the maxima to become more  precisely defined, in the limit of a clean sample and small $\Delta B$.

Experimentally, the values of $\sigma_{xx}(q, \omega)$, at relatively low frequencies, have been  extracted from accurate measurements of the propagation velocity of surface acoustic waves, as a function of acoustic wavelength  and applied magnetic field, in a sample containing a two-dimensional electron gas, by Willett and coworkers\cite{willettSAW93}.
  These surface acoustic wave experiments  were, in fact,  very important in establishing the validity of the HLR picture.

Another type of commensurability oscillation, commonly referred to as {\em Weiss oscillations},  may be observed by measuring the dc resistivity in the presence of a periodic electrostatic potential, which may be imposed by a periodic array of gates or etched defects on the surface\cite{kamburov14,kang93,smet96,smet98,smet99, willett99,zwerschke99}.  In this case, theory predicts, and experiments have seen,  maxima in the resistivity at magnetic fields where the wave vector $q$ of the array satisfies approximately Eq (\ref {qn1}) or (\ref{qn2}).

In the following subsections, we shall examine a third type of commensurability oscillation related to the existence of local minima in the spectrum $\omega(q)$ of so-called magnetoroton excitations in a fractional quantized Hall state with $\nu$ close to 1/2.  Magnetorotons may be understood as  bound states of a quasiparticle in the lowest empty composite-Fermion Landau level and a quasihole in the highest filled level.  As was discussed by Simon and Halperin\cite{SimonHalperin}, the spectrum should have a series of minima, at wave vectors given approximately by Eq. (\ref{qn1}), which become increasingly sharp for small values of $|\Delta B|$.  The frequencies $\omega (q)$ are manifest as poles in the response function to an applied  electric field at frequency $\omega$ and wave vector $q$.  For certain filling fractions the magnetoroton minima have been numerically calculated using composite fermion trial wave functions\cite{jainroton}.

Although the magnetoroton spectrum may be difficult to measure experimentally in the region of interest to us{\footnote{However, the magnetoroton spectrum has been successfully measured at filling fractions $2/5$, $3/7$ and $4/9$ by Kukushkin et al.\cite{Kukushkin}.}}, it has a big advantage from a theoretical point of view compared to predictions for the magnetoresistance in a periodic potential or the zero-frequency longitudinal conductance.  The last two quantities are well defined only in the presence of a small but finite density of impurities.  However, the behavior of a partially full composite-fermion Landau level in the presence of weak impurity scattering may be quite complicated, and is certainly not well understood.
By contrast, the magnetoroton spectrum may be studied in system without impurities, in a fractional quantized Hall state where there is an energy gap and where the magnetoroton may be precisely  defined, as  the lowest energy excitation for the given value of $\bfq$. We shall comment briefly on our understanding of the Weiss oscillations at the end of this section.

The requirements imposed by PH symmetry on the magnetoroton minima were stated in the Introduction. They are not satisfied in a naive application of the HLR approach. Below we show how they are satisfied by a more careful application of the HLR theory.

\subsection {Magnetoroton spectrum at $\nu = p / (2p+1)$}

 We now look for the dispersion minima of the magneto-roton modes within the HLR composite fermion theory, at filling fraction $\nu=\frac{p}{2p+1}$, when $|p|$ is large.   The magnetoroton frequencies will appear as poles in the current response matrix
 $\hat{W} (\bfq, \omega)$ to an electric field ${\bf{E}}$ at wave vector $\bfq$ and frequency $\omega$, defined by
\be
\bfj (\bfq, \omega)  = \hat{W} \,  {\bf{E}} (\bfq, \omega).
\ee
 We shall take $\bfq$ to lie along the $x$-axis, so the indices $x$ and $y$ refer to longitudinal and transverse components respectively.

Our analysis will follow closely the work of SH\cite{SimonHalperin}, and we shall first consider the response function using the RPA.  Following Eqs. (27) and (28) of SH, we may write
\be
\label{Wtensor}
\hat{W}^{-1} = \hat{\rho} + \hat{U},
\ee
\be
 \hat{\rho} = \hat{\rho}^\cf (\bfq, \omega)+ \hat{\rho}^{\cs} ,
 \ee
 where $\hat{\rho}^\cf (\bfq, \omega)= (\hat{\sigma}^\cf)^{-1} (\bfq, \omega)$ is the resistivity tensor of the composite fermions, and $\hat{U}$ has matrix elements
 \be
 U_{xx} =  i \frac {q^2}{\omega} v_2(q) , \,\,\,\, U_{xy} = U_{yx} = U_{yy} = 0 ,
 \ee
where $v_2$ is the two-body  interaction, defined above.

According to SH, the composite fermion conductivity tensor, for a general value of $p$, can be expressed in terms of an infinite sum  of terms involving associated Laguerre polynomials.  It the limit of large $p$, one can employ a semiclassical approximation, where the sums can be carried out, and one can write  the conductivity tensor in closed form in terms of Bessel functions. For the moment, we shall employ this semiclassical approximation, and shall comment later on the corrections that would be expected if one were to employ the full expressions for   $\hat{\sigma}^\cf (\bfq, \omega)$.

\subsubsection {Semiclassical calculation of $\hat{\rho}$}
\label{rhozeros}

The semiclassical results of SH may be written
 (in units where $e^2/h=1/2\pi$)  as
\bea
\label{sigmaCF}
\sigma^{\cf}_{xx}&=&i\frac{2pR}{\pi X^2}\left[-\frac{1}{2}+\frac{\pi R}{2{\rm sin}(\pi R)}J_R(X)J_{-R}(X)   \right], \nn
\sigma^{\cf}_{xy}&=&i\sigma^{\cf}_{xx}+\frac{ pR}{X{\rm {sin}}(\pi R)}J_{R+1}(X)J_{-R}(X), \nn
\sigma^{\cf}_{yy}&=&\sigma^{\cf}_{xx}+i\frac{ p}{{\rm sin}(\pi R)}J_{R+1}(X)J_{1-R}(X),
\eea
where $R \equiv \omega/\Delta\omega_c$ and $X \equiv qR^*_C=\frac{|2p+1| qk_F}{B}=\frac{2|p|q}{k_F}$ ($R^*_C$ is the cyclotron radius of the composite fermion), $\Delta \omega_c = \Delta B / m$,
and $J_{\nu}(X)$ is the Bessel function of the first kind. The full resistivity is given by the composition rule
\be
\rho=(\sigma^{\cf})^{-1}-4\pi\hat{\epsilon}.
\ee

We begin by looking for  the poles of the physical conductivity tensor, which correspond to zeros of ${\rm Det}(\rho)$. To leading order in $1/p$, these poles are located at  the zeros of  ${\rm Det}(\sigma^{\cf})$, which would yield dispersion minima at $X=z_n, R=0$ where $z_n$ is the $n$'th zero of the Bessel function $J_1$. Here, however, we calculate the momenta ($\propto X$) at these dispersion minima to next order in $1/p$ and address the question of  their PH symmetry near half-filling.

To leading order in $R$ and $\Delta X=X-z_n$, the $\sigma^{\cf}$ tensor is given by
\bea
\label{sigmaCFleading}
\sigma^{\cf}_{xx}&=&i\frac{J_0^2(z_n)-1}{\pi z_n^2}pR, \nn
\sigma^{\cf}_{xy}&=&\frac{J_0^2(z_n)}{\pi z_n}p\Delta X, \nn
\sigma^{\cf}_{yy}&=&i\frac{J_0^2(z_n)-1}{\pi z_n^2J_0^2(z_n)}pR+i\frac{J_0^2(z_n)}{\pi}\frac{p(\Delta X)^2}{R},
\eea
where the following Bessel function identities were used to reach the above result:
\bea
J_0(z)&=&J_1'(z)+\frac{J_1(z)}{z}, \nn
\partial_{\nu}J_{\nu=1}(z)&=&\frac{\pi}{2}Y_1(z)+\frac{J_0(z)}{z}, \nn
\frac{2}{\pi z}&=&J_1(z)Y_0(z)-J_0(z)Y_1(z).
\eea
We are looking for values of $R$ and $\Delta X$ that satisfy
\be
\label{mini}
{\rm Det} ( \hat{\rho} \hat{\sigma} ^{\cf})={\rm Det} (1-4\pi\hat{\epsilon}\sigma^{\cf})=0.
\ee
Using Eq.~\eqref{sigmaCFleading}, we find the dispersion curve
\begin{widetext}
\be
\label{dispersion}
\left[\frac{4(J^2_0(z_n)-1)}{z_n^2J_0(z_n)}\right]^2(pR)^2=\left[\frac{4J_0(z_n)}{z_n}\right]^2\left(p\Delta X+\frac{z_n}{4}\right)^2+(1-J^2_0(z_n)).
\ee
\end{widetext}

The dispersion minima are then given by
\be
\Delta X=-\frac{z_n}{4p},
\ee
which  means that at $\nu=p/(2p+1)$, we have
\be
X= z_n\left(1-\frac{1}{4p}\right).
\ee
Since the composite fermion Fermi momentum $k_F$ is determined solely by the electron density in the HLR theory, we have
\be
X= \frac{2|p|q}{\sqrt{\frac{2pB}{2p+1}}}\sim \frac{2q|p|}{\sqrt{B}}\left(1+\frac{1}{4p}\right),
\ee
which gives
\be
\label{mrminima}
q_n\sim\frac{z_n\sqrt{B}}{2|p|}\left(1-\frac{1}{2p}\right).
\ee
For $p=p_0$ with $p_0$ positive, we have $\Delta B = B/(2p_0+1)$ and
\be
q_n\sim\frac{z_n\sqrt{B}}{2p_0}\left(1-\frac{1}{2p_0}\right) \sim \frac{z_n \Delta B}{B^{1/2}} ,
\ee
while for $p=-p_0-1$, we have $\Delta B = -B/(2p_0+1)$ and
\be
q_n\sim\frac{z_n\sqrt{B}}{2(p_0+1)}\left(1+\frac{1}{2p_0}\right)\sim\frac{z_n\sqrt{B}}{2p_0}\left(1-\frac{1}{2p_0}\right),
\ee
which is again equal to $z_n |\Delta B|/B^{1/2}$. This is consistent with PH symmetry, at least to order $1/p^2$.

The frequencies corresponding to these dispersion minima are given by
\be
\label{omegasc}
\omega_n=\frac{z_n^2J_0(z_n)}{4  |p|  \sqrt{1-J_0^2(z_n)}}  |\Delta\omega_c| \,  .
\ee

As we will see below, the exact values of $q_n$ and $\omega_n$ will receive significant corrections once we take other effects into account. However, particle-hole symmetry of the dispersion will still hold even after we include all the leading corrections.

\subsubsection {Corrections for the poles of $\hat {W} $}

We now discuss various corrections to the above result. The regime we are interested in, for $p\gg 1$, will have $\Delta X \sim 1/p$ and $R\sim1/p^{1/2}$. In this regime, the components of   $\hat{\sigma}^{\cf}$  in Eq.~\eqref{sigmaCFleading} will be of order $p^{1/2}$ or $p^0$, and any correction of higher order in $1/p$ will not affect our results.

First we consider Fermi-liquid corrections including mass renormalization and the residual Landau interaction. To incorporate mass renormalization we simply replace $\Delta \omega_c$ by $\Delta \omega^*_c=\Delta B  /m^*$. This leads to a violation of Kohn's theorem and the $f$-sum rule, which has to be compensated by introducing the proper Landau interaction parameter $F_1$. The Landau parameter leads to another contribution to the diagonal components of the composite fermion resistivity tensor, $\Delta\rho^{\cf}_{xx} =\frac{i(m^*-m)\omega}{n_{\rm{el}}   } $, which is of order $1/p^{3/2}$ in the regime we consider. This will not change our result for the dispersion minima.

We can also consider corrections to the semiclassical expression of $\sigma^{\cf}$ in Eq.~\eqref{sigmaCFleading}, for example from the full quantum-mechanical summation in Appendix A of SH.\cite{SimonHalperin} 
Since we expect the semiclassical expression to be justified in the large $p$ limit (which has been explicitly demonstrated recently in \cite{Gromov}),   the corrections should be formally higher order in $1/p$. In principle several leading order corrections are possible:
\bea
\label{RPAcorrection}
\Delta\sigma^{\cf}_{xy}&=&a\frac{p}{|p|}, \nonumber \\
\Delta\sigma^{\cf}_{yy}&=&    b\frac{ip}{|p|R} +    c \frac{ip\Delta X}{|p|R}+d \frac{i}{pR},
\eea
where other types of corrections are either higher order in $1/p$ (taking into account $\Delta X\sim R\sim 1/p$), or forbidden by general constraints. These constraints include $\sigma^{\cf}$ being odd under $p\to-p$ when fixing $R$, and $\sigma^{\cf}_{xx},\sigma^{\cf}_{yy}$ being odd under $R\to-R$ when fixing $p$. Both constraints are closely related to the symmetry of the conductivity matrix elements under a change of the sign of the frequency.
These terms would give rise to  corrections to the dispersion curve in Eq.~\eqref{dispersion}, which would lead to corrections of the locations of the dispersion minima, so that
\be
\label{alph1}
X_n=z_n\left(1-\frac{1}{4p}+\frac{\alpha}{|p|} \right),
\ee
with some constant $\alpha$. The actual momenta at the dispersion minima would thus be  shifted to
\be
\label{momentacorrection}
q_n\sim \frac{z_n\sqrt{B}}{2|p|}\left(1-\frac{1}{2p}+\frac{\alpha}{|p|}\right).
\ee
These corrections beyond the semiclassical approximation could indeed shift the momenta of the magneto-roton minima at order $1/p^2$. However, this correction would be symmetric in $p\to-p-1$ (at order $1/p^2$), so particle-hole symmetry is still preserved at this order.

The correction terms in (\ref{RPAcorrection})  will also lead to a correction of the frequencies at the minima:
\be
\label{omegacorrection}
\Delta(\omega_n^2)_{\rm{Quantum}} \sim \frac{\beta}{p}(\Delta\omega_c^*)^2,
\ee
with some constant $\beta$.  This gives a frequency  $\omega_n$ of order $ p^{-1/2 }\omega_c^*$ , which is parametrically larger than  the semiclassical result in Eq.~\eqref{omegasc}.

We have calculated numerically the values of the coefficients $a,b,c$ and $d$ in Eq. (\ref{RPAcorrection}) at the first two magnetoroton minima, $n=1, 2$. We find that the coefficients $a, b$, and $c$ are all zero, and consequently, $\alpha=0$ in Eqs. (\ref{alph1}) and(\ref{momentacorrection}). The values of $d$, are nonzero, however, being equal to 0.082 at $n=1$ and  and 0.297 at $n=2$. These lead to values of $\beta$ equal to 0.0046 and 0.029, respectively, in Eq. (\ref{omegacorrection}).   Hence,  corrections due to the difference between the semiclassical expressions in terms of Bessel functions, and the full quantum sum in SH can  affect the frequency  at the magnetoroton minimum, but do not actually contribute a shift in the wave vectors, to order $|\Delta B|^2$.

Finally we notice that the real dispersion curve is given by the poles of the full response tensor $\hat{W}$ in Eq.~\eqref{Wtensor}. This modifies Eq.~\eqref{mini} to
\be
{\rm Det} (\hat{W}^{-1}\sigma^{\cf})={\rm Det} (1-4\pi\hat{\epsilon}\sigma^{\cf}+\hat{U}\sigma^{\cf})=0.
\ee
This leads to an extra term
\be
U_{xx}\sigma^{\cf}_{xx}=\frac{(1-J_0^2(z_n)) | p|  q^2v_2(q)}{\pi z_n^2  |\Delta\omega_c^*|  }
\ee
on the right hand side of the dispersion relation in Eq.~\eqref{dispersion}. Generically this term is dominating over the other terms in Eq.~\eqref{dispersion}. To see this, let us consider very long-ranged interaction $v_2(q)\sim \frac{1}{|q|^{1+\delta}}$, which gives rise to simple Fermi-liquid behavior at low energy. In this case the above term becomes
\bea
\label{extraterm}
U_{xx}\sigma^{\cf}_{xx}&\sim& \frac{p|q|^{1-\delta}}{\Delta\omega^*_c}\sim |p|^{1+\delta}X^{1-\delta} \nonumber \\
&\sim& z_n^{1-\delta}|p|^{1+\delta}+(1-\delta)z_n^{-\delta}|p|^{1+\delta}\Delta X,
\eea
where we have used the fact that $\Delta\omega_c^*\sim 1/p$. The first term $\sim|p|^{1+\delta}$ dominates over the other terms in the original dispersion curve Eq.~\eqref{dispersion}. Its effect is to set the frequency at the dispersion minima, in leading order, to be
\be
\label{domeg}
\omega_n=\frac{z_nJ_0(z_n)}{4|p|}\sqrt{\frac{p\Delta\omega_c^*q_n^2v_2(q_n)}{\pi(1-J_0^2(z_n))}}.
\ee

In the physical case of Coulomb repulsion, $v_2(q)=2\pi/\epsilon q$ where $\epsilon$ is the dielectric constant, (\ref{domeg})  still gives the leading result for the minimum frequency, but one should  take into account the variation of  $\Delta \omega^*$ due to logarithmic divergence of the effective mass.  Specifically it is predicted that  \cite{sternSH95,morfMAS02}  
\be
\label{Domcoulomb}
|\Delta \omega^*| =\frac{|\Delta B|}{m^*} \sim \frac{ \pi  e^2} { 2 \epsilon l_B |2p+1| [C+ \ln |2p+1| ]},
\ee
where the constant $C$ depends on the bare mass  and on the behavior of the interaction at short distances.  For pure Coulomb interactions and vanishing bare mass, the best available estimate is  $C \approx 4.1$\cite{morfMAS02}.

The second term in Eq.~\eqref{extraterm} leads to a shift in the momenta at the minima, giving
\be
\label{momentacorrection2}
q_n\sim \frac{z_n\sqrt{B}}{2|p|}\left(1-\frac{1}{2p}-\frac{\gamma}{|p|^{1-\delta}}\right).
\ee
The extra shift is parametrically dominating, but it does not depend on the sign of $p$, so it does not affect particle-hole symmetry, at least to the order $|p|^{-2} $ that we are considering. For Coulomb repulsion the correction is of the form $\sim{\rm log}|p|/|p|$, which is again particle-hole symmetric.

The predicted magnetoroton spectrum for the symmetric fractions $\nu=20/41$ and $\nu=21/41$ are plotted in Figure 1, at our various levels of approximation, for the case of pure Coulomb interactions.

\begin{center}
\begin{figure}[ht]
\captionsetup{justification=raggedright}

\adjustbox{trim={0\width} {.1\height} {0\width} {.1\height},clip}
{\includegraphics[width=1\columnwidth]{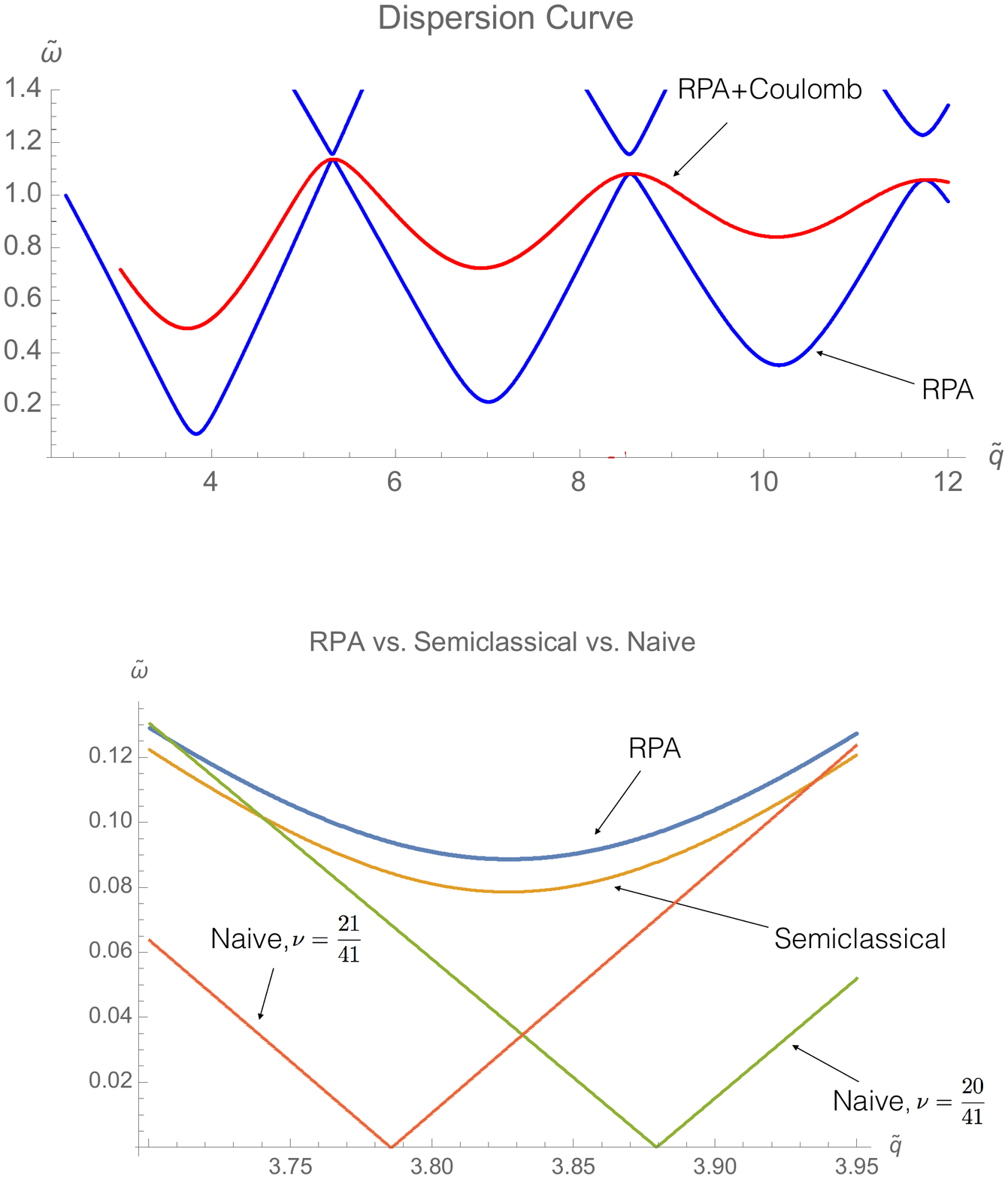}}
\label{figure1}

\caption{
Magnetoroton spectrum at fractions $\nu=20/41$ and $\nu=21/41$.  Plots show the 
reduced frequency $\tilde {\omega} \equiv \omega / |\Delta \omega^*|$ versus  the reduced wave vector
$\hat{q} \equiv q l_B / |2 \nu - 1 |$.
The curve labeled ``RPA + Coulomb" shows the  
magnetoroton spectrum  computed in the HLR approach, including the correction due to the Coulomb interaction. 
The curves labeled ``RPA"  and ``Semiclassical" show  the locations of the  poles in the electron conductivity tensor $\hat{\sigma} (\bfq,\omega)$, which does not include the interaction effect,   computed in  the Random Phase Approximation  and semiclassical approximation, respectively. Curves for $\nu=20/41$ and $\nu=21/41$ could not be distinguished in these plots.  The expanded figure in the lower panel includes for comparison  a naive    approximation, which identifies the magnetoroton spectrum with the zeros of the determinant of the composite fermion conductivity $\hat {\sigma}^\cf (\bfq, \omega)$. Although the naive
 approximation coincides with the RPA and semiclassical approximations to leading order in the deviation
from $\nu=1/2$, it deviates from them at second order and is not symmetric about $\nu=1/2$ at this order.  
}
\end{figure}
\end{center}

\subsection {Magnetorotons near $\nu=1/(2N)$}

The  analysis given above can be readily extended to the magnetoroton spectrum in fractional quantized Hall states of the form
\be
\nu =  \frac {p}{2pN+1} ,
\ee
where $N$ is an integer  $>1$, which are close to $\nu=1/(2N)$, for large $|p|$.   Here we define $\Delta B$ as 
\be
\Delta B  \equiv B - 4 \pi N n_{\rm{el}}  = \frac {B}{2pN+1},
\ee
which is the difference between $B$ and the value corresponding to $\nu=1/(2N)$ at the given electron density.
Using the same analysis as for $N=1$, we find that the  minima of the lowest magnetoroton modes occur at momentum values $q_n$ which depend on the absolute value,  but not on the sign, of $\Delta B$, at least through order $|\Delta B|^2$, provided we compare systems with different electron densities but the same magnetic field $B$. Specifically, we have 
\be
q_n = \frac { z_n |e \Delta B| l_B }{\hbar N^{1/2} },
\ee
up to  small corrections which are symmetric in $\Delta B$. 
 Along with our previous result that in the presence of PH-symmetric disorder, the Hall conductance at $\nu=1/(2N)$ is fixed at $1/(4 \pi N)$, at least through second order in the impurity scattering rate, this suggests that there is a type of emergent particle-hole symmetry near all these even-denominator fillings.

Interestingly, a similar type of emergent particle-hole symmetry was found also when the energy gaps  $E_{\rm{G}}$ of fractional quantum Hall states at filling factors close to $1/(2N)$ were calculated for electrons interacting through the Coulomb interaction. The energy gap, in this case, is predicted to have the form  \cite{sternSH95} 
\be
\label{Dom2mcoulomb}
|\Delta \omega^*| \sim \frac{ \pi  e^2} { 2 \epsilon l_B N^{3/2}   |2pN+1| [C'+ \ln |2pN+1| ]},
\ee
which reduces to Eq.~(\ref{Domcoulomb}) for $N=1$.
This expression is predicted to be exact in the limit of large $p$, and the leading logarithmic term is independent of the bare mass of the electron electron.  Moreover, the result is symmetric in $\Delta B$, at least to lowest order.  However, the possibility of asymmetric corrections at second order in  $\Delta B$ was not investigated.


\subsection {Weiss oscillations}

As remarked above, a proper analysis of the experiments measuring the resistivity in the presence of an imposed periodic potential with wave vector $\bfq$  would require a careful analysis of the effects of impurity scattering at filling factors away from $\nu=1/2$, which is beyond the scope of the current paper. 
However, one can gain insight into the problem from a very recent investigation by Cheung, 
Raghu and Mulligan (\cite{cheungCRM06} and private communications).  They have calculated the change in resistivity $\Delta \rho_{xx}$ produced by a weak modulating potential in an approximation where they treat impurities in a simple relaxation approximation, 
where the relaxation rate is take to be a constant, independent of $\Delta B$ and the  scattering wave vector, etc. Although the bulk of their paper is based on the Son-Dirac model, they also present results  based on the HLR equations.  

Treating the ratio $x = V(\bfr) m^* / \delta b(\bfr)$ as a free parameter, where $V$ is the residual screened electric potential  produced by the external periodic potential and $\delta b$ is the induced Chern-Simons magnetic field seen by the composite fermions, they find a series of curves for the induced magnetoresistance, as a function of $\Delta B$  whose shapes depend on $x$. When $x=1/2$, they find that the HLR prediction coincides precisely with the Dirac prediction and  is properly symmetric in $\Delta B$, when the density is varied while $B$ is held fixed. In particular, when $x=1/2$, it is predicted that there will be minima in 
$\Delta \rho_{xx}$ at magnetic fields that satisfy   
\be
\label{dbn}
|\Delta B| \sim \left( \frac{ B \hbar}{e}\right)^{1/2}  \frac{q}{z_n} .
\ee
According to the discussion in Subsection \ref{sec:Dpm}  of the present paper, leading to Eq.~(\ref{bV}), the value $x=1/2$ is indeed the proper choice for that parameter.  (We note that the Weiss oscillations are measured at a temperature $T$ that is larger than the energy scale $|\Delta B|/m^*$, so that the electron compressibility may be taken to be the same as at $\nu=1/2$.)

The fact that one must take into account modulations in the Chern-Simons scalar potential as well as in the Chern-Simons magnetic field, in order to understand in a quantitative way  the effects of an imposed periodic potential on the electrical resistivity, was previously emphasized by Zwerschke and Gerhardts \cite{zwerschke99}. Also, a correct formula for the magnetoresistance in the presence of modulations in both the screened electrostatic potential and the effective magnetic field $b$ was contained in Ref.~\cite{Barkeshli15} by Barkeshli, Mulligan, and Fisher. In that paper, however, authors then ignored the electrostatic potential on the grounds that its effects would be small compared to the effect of $b$, so they  did not  obtain  the small correction   necessary to restore the PH symmetry.

Although the resistance minima observed experimentally in Ref.~\cite{kamburov14} do obey particle-hole symmetry, the actual positions deviate (symmetrically)  from the values predicted by Eq.~(\ref{dbn}), by amounts of order  $|\Delta B|^2$.  We do not know whether these deviations could be explained by a theory that includes the effects of impurity scattering in a more accurate way.

It should be emphasized that theoretical discussions about presence or absence of particle-hole symmetry generally refer to a situation where $n_{\rm{el}}$ is varied while $B$ is held constant.  In experiments, however, it is most common to vary $B$ while $n_{\rm{el}}$ is held constant.  In that mode of operation, features that occur at positions we consider symmetric, such as  those given by Eq. (\ref{dbn}), will appear asymmetric in the data, by amounts of  order $|\Delta B|^2$.  By contrast, the values of $|\Delta B|$ given by the naive HLR theory, where modulations in the Chern-Simons  scalar potential are ignored, would appear symmetric 
about $\nu=1/2$ in the data.  

{
\subsection{Ambiguity of $k_F$}

The question of what determines the Fermi momentum $k_F$ of composite fermions, away from half-filling, has played a significant role in the literature\cite{kamburov14,Barkeshli15,JainLuttinger}. Naively, there are three possible answers depending on which theory one uses: in HLR theory the Fermi volume is given by the particle density, in anti-HLR it is given by the hole density, and in Son-Dirac it is given by the half of the flux density. These answers are identical at $\nu=1/2$, but deviate from one another away from half-filling. However, one should be more careful when addressing this issue. 

There are two sources of confusion regarding $k_F$. First, $k_F$ of the composite fermions is not a sharply defined quantity away from half-filling, since the composite fermions move in a nonzero effective magnetic field $\Delta B$ and do not have a sharp Fermi surface. The ambiguity in the definition of $k_F$, set by the inverse effective cyclotron radius, is of order $\Delta B$. The differences in $k_F$ determined from electron, hole or flux densities are also of this order, so the three answers are identical within this intrinsic ambiguity.

A subtler point is that $k_F$ itself is not a measurable quantity, especially away from $\nu=1/2$. What can be measured in commensurability oscillation experiments are the commensurability momenta $q_n$. Past work has inferred $k_F$ from  $q_n$ via Eq.~\eqref{qn1}. However, the simple relation Eq.~\eqref{qn1} is valid only to leading order in $\Delta B$. Once we go to higher order in $\Delta B$, which is necessary to differentiate particle-density from hole-density, the simple relation Eq.~\eqref{qn1} no longer holds and a more careful calculation is needed. This is exactly what we did in the earlier parts of this Section. Our results show that the commensurability momenta are indeed particle-hole symmetric, even though in HLR theory $k_F$, which is not an observable by itself, appears to be formally PH asymmetric.}


\section {Comparison with the Dirac theory\label{sec:dirac}}

The Son-Dirac model may be defined by Lagrangian density of the form
\begin{eqnarray}
\label{dirac}
\mathcal{L}_D = &\psi^\dagger (i D_t - \mu - i v_D \, {\bf{D}} \times {\bf{\sigma}} - m_D \, \sigma^z) \psi  +
\nonumber  \\
&+ \left[ \frac {AdA} {8 \pi }   +  \frac{adA}{4 \pi } - \frac {ada} {8 \pi } \frac{m_D}{|m_D|}  \right]  +
 \mathcal{L}_{\rm {int}} ,
\end{eqnarray}
\be
D_\mu \equiv \partial_\mu + i \, a_\mu ,
\ee
where  $\psi$ is a two-component Grassmann spinor, $\bfA$ is the external  magnetic  field, $\mathbf{\sigma}$ are the Pauli spin matrices, and $ \mathcal{L}_{\rm {int}} $ is a term which represents the two-body interaction $v_2$.  The velocity $v_D$ is an input parameter, like the effective mass $m^*$ in the HLR theory, which must be taken either from experiment or from an independent microscopic calculation.  We shall be interested in a situation in which the Fermi level is inside the band of positive energy fermion states. The lower Dirac band is integrated out, which produces the $\pm\frac{1}{8\pi}ada$ term.

The Son-Dirac Lagrangian becomes explicitly PH symmetric if one takes the limit $m_D \to 0$. In this limit, the contribution of the $ada$ term is  precisely canceled by  the contribution from the Berry curvature, which is completely concentrated  at the bottom of the occupied states in the positive energy Dirac band.   Then, the Lagrangian may be replaced by a form in which $m_D$ is precisely zero and the $ada$ term is simply omitted; {\it i.e.}, there is no longer a Chern-Simons term in the action for the gauge field $a_\mu$   In the following discussion, we confine ourselves to the case $m_D$=0, except where otherwise specified.

In the Son-Dirac formulation, the composite fermions see an effective magnetic field  $b(\bfr)$ which is related to the electron density and the applied magnetic field in the same way as in HLR:
\be
b =  \nabla \times \bfa =  4 \pi n_{\rm{el}} - \nabla \times \bfA .
\ee
However, the electron density and the composite fermion density are no longer identical.  Rather, the density of Dirac composite fermions   is tied to the (local) value of the magnetic field
\be
n_\df = : \psi^\dagger \psi: =  - \frac {1}{4 \pi} \nabla \times \bfA .
\ee
Similarly, the current of the Dirac fermions is related to the   local electric field  by
\be
\bfj_\df =  - \frac {1}{4 \pi} \hat{z} \times  \bf{E},
\ee
while the  effective electric field felt by the Dirac fermions is given by
\be
\bfe_\df = -\nabla a_0 - \partial_t \bfa =  \hat{z} \times (4 \pi  \bfj_{\rm{el}} - \bf{E} )  .
\ee
The electrical conductivity tensor, for a long-wavelength electric field is then given by
\be
\label{sigrhodf}
\hat{\sigma} = \hat{\rho}^\df + \hat{\sigma}^\cs ,
\ee
where $\hat{\rho}^\df  = (\hat{\sigma}^\df)^{-1} $ is the resistivity tensor of the Dirac fermions, and
\be
\hat{\sigma}^\cs = \frac{1}{4 \pi} \hat{\varepsilon} .
\ee

As in  the HLR theory  the presence of potential disorder will cause fluctuations in electron density, which will lead to fluctuations in the effective field $b(\bfr)$  proportional to  the self-consistent electric potential $V(\bfr)$. Potential fluctuations do not lead to fluctuations in $a_0$ or in the effective electric field $\bf{e}$.  Therefore, if the potential fluctuations are statistically PH symmetric, so that all odd moments of $b$ are zero, the Dirac fermions will see a field that is statistically time-reversal symmetric, and $\hat{\rho}^\df $ will be purely diagonal.  Therefore, we recover $\sigma_{xy} = 1/ 4 \pi$ as required by particle hole symmetry.

At a finite frequency $\omega$, in the absence of impurities, in the RPA, the resistivity tensor for Dirac fermions is readily calculated to be
\be
\hat{\rho}^\df  = - i \omega m^* / n_{\rm{el}}
\ee
where $m^* = k_F / v_D$.  As this is purely diagonal, the ac Hall conductivity remains fixed at the value required by PH symmetry.  However, the diagonal conductivity $\sigma_{xx}$   predicted by (\ref{sigrhodf})  does not agree with the  result $\sigma_{xx}(\omega) = 0$, which is required by Kohn's theorem in the limit where the electron mass $m \to 0$, and electrons are restricted to the lowest Landau level. As remarked above, this can be corrected, beyond the RPA,  by including the effects of Landau interaction parameters
$F_{\pm 1}$.

Using the Son-Dirac Lagrangian for $m_D=0$, one predicts that fractional quantized Hall states should occur when
\be
\Delta B = \frac {B}{2 p_\df},
\ee
where $p_\df$ is half of an odd integer, either positive or negative.  This condition is obviously PH symmetric and it is equivalent to the HLR prediction, with the identification $p_\df = p + 1/2$. The shift in the choice of indexing reflects the presence of a Berry phase of $\pi$ for the Dirac fermions at the Fermi energy.  The energy gaps in the quantized Hall states are given, within RPA by Eq. (\ref{Eg}) with $m$ replaced by $m^* = k_F / v_D$.  As remarked previously, the gaps will obey PH symmetry provided that the velocity  $v_D$ is assumed to depend on the magnetic field, and not on the electron density, or more generally, if $v_D$ is assumed to be an even function of $\Delta B$.

According to PH symmetry,  the magneto-exciton spectra should  also be independent of the sign of $\Delta B$.  The positions of the magnetoroton minima may be found, to lowest order in $\Delta B$, by
tracking the dispersion of poles in the electrical conductivity $\hat{\sigma}(q, \omega)$, as was done in Subsection \ref{rhozeros} above in the HLR picture.  Taking into account Eq.~(\ref {sigrhodf}), we see that within the Dirac description, poles in $\hat{\sigma}(q, \omega)$ coincide with the occurrence of a zero in the determinant of the composite fermion conductivity tensor $\hat{\sigma}^\df (q, \omega)$.  Within the semiclassical approximation, these zeros  occur at $\omega=0$, if
\be
q_n = z_n \frac { B^{1/2} } {2  |p_\df |}.
\ee
These values are  clearly PH symmetric and are identical to the results obtained using HLR in Subsection
\ref{rhozeros}, through order $| \Delta B |^2$. This result for Dirac composite fermions was also obtained in \cite{sonroton}, to lowest order in $|\Delta B|$, with careful attention to interaction effects.

As in the HLR case, the actual locations of the magnetoroton minima  in the Dirac theory will be shifted from these values (by amounts small compared to $q_n$), and the frequency values will be shifted from zero,  due to interaction effects and to corrections to the semiclassical theory, but
all such shifts should be symmetric in $\Delta B$.

Finally, we discuss  properties of the Dirac Lagrangian  (\ref{dirac})  in the case where the Dirac mass $m_D$ is not set equal to zero, so the theory  is not explicitly PH symmetric.  As was observed by Son\cite{sonphcfl},  in the non-relativistic limit, where $m_D v_D \gg k_F$, the Dirac action  reduces precisely to the  HLR action
(\ref{Lhlr}), after a redefinition of  the  gauge field, ($a_\mu \to a_\mu + A_\mu$). As we have seen, the HLR theory and the massless Dirac theory give identical results for long-wavelength low-energy properties in the limit of $\nu=1/2$, so that PH symmetry reappears in this case. We find that there is a similar  emergent PH symmetry for intermediate values of $m_D$.   Since the Lagrangian for the Dirac theory with finite $m_D$ includes a Chern-Simons term identical to that in the HLR theory, the relations between the composite fermion and the electronic  response functions are identical in the two theories.  The semiclassical theory for the minima of the magnetoroton spectra take the same form as we found in Section \ref{sec:commens} above, which implies that the spectrum is, again, symmetric in $\Delta B$, at least through order $|\Delta B|^2$.
Similarly, we find that the Hall conductance in the presence of impurities at $\nu=1/2$ is fixed at $1/4 \pi$, at least through order $(1 / l_{\cf})^2$, under the same conditions that we assumed  in the analysis of HLR in Section \ref{sec:dc}.  

An apparent difference between HLR and a Dirac theory with finite $m_D$ is that in the latter case  the fermions near the Fermi energy  have a non-zero Berry curvature. This Berry curvature is the same as that which results from spin-orbit coupling in a semiconductor, 
which, as we have remarked, is responsible for side-jump contributions to the anomalous Hall effect in semiconductor models. However, in the limit of scattering wavevectors $q$ much smaller than $k_F$, which we have assumed in our analysis, the matrix element for the spin-orbit term is negligible compared to that  from  the screened impurity potential $V$ or the effective magnetic field fluctuation $b$.  Scattering from potential fluctuations with $q$ of order $k_F$ would depend on renormalized matrix elements whose values are beyond the scope of an effective theory.

In the Dirac theory with finite $m_D$, fermions at the Fermi energy will have a Berry phase which is neither zero nor $\pi$.  In contrast with the Berry curvature, the total Berry phase has no direct effect on the dc Hall conductivity in the presence of impurities, but it does affect the ac Hall conductivity. Just as in HLR, the finite frequency Hall conductivity will deviate from $1/4\pi$ at order $\omega^2$,  unless the effect is counteracted by a non-zero Fermi-liquid interaction parameter, whose actual value will depend on details of the original microscopic theory.

\section{Conclusions}

We have seen that in the limit of long wavelengths and low frequencies, with $\nu$ close to 1/2, and in the limit of small disorder potential,  the Son-Dirac and HLR theories make identical physical predictions for several key properties, provided that the HLR theory is properly evaluated.  Both theories give results for these properties that are consistent with PH symmetry, even at the RPA level.  In the Dirac theory,  PH symmetry is put in by hand, at the outset, by setting the Dirac mass $m_D$ equal to zero.  In the HLR theory, PH symmetry seems to emerge, asymptotically, in this limit, even though it is not put in at the beginning.  Moreover, the PH symmetry seems to emerge even if the bare mass $m$ is not taken to zero, which would be the condition for electrons to be confined to a single Landau level, where PH symmetry would be exact.

In order to get the correct energy scale for the specific heat or for energy gaps in fractional quantized Hall states close to $\nu=1/2$, at the RPA level, the bare mass $m$  in HLR must be replaced by a renormalized mass $m^*$, whose value cannot be obtained within the theory itself.  Similarly in the massless Dirac theory, one must use a renormalized value of the Fermi velocity $v_D$. After these substitutions are made, however, neither the Dirac theory nor the HLR theory will give the correct response functions to perturbations at a finite frequency, unless one also includes the  effects of the Landau interaction parameters $F_l$, for $l = \pm 1$. In the HLR theory, this correction gives the correct frequency response, dictated by the Galilean invariance of the original model. In the limit $m \to 0$, this leads to a conductivity tensor
$\hat {\sigma} (\omega) $ for a spatially uniform electric field that  is independent of  $\omega $ and which, therefore, satisfies the requirement  that $\sigma_{xy} (\omega)$ should be independent of frequency by PH symmetry, for  electrons confined to the lowest Landau level. If the Landau interaction were omitted,  however, an RPA calculation with the renormalized mass would incorrectly give a frequency-dependence to $\hat{\sigma}$, which would result in  a non-zero correction to $\sigma_{xy} (\omega)$ at order $\omega^2$.

In the Dirac theory, for $m_D = 0$, one obtains correctly $\sigma_{xy} (\omega) = 1/ 4 \pi$ at all frequencies, even at the simple RPA level, because of the explicit built in PH symmetry.  However, the diagonal conductance $\sigma_{xx} (\omega)$ will  be incorrect at order $\omega$, unless one includes the Landau interaction correction.

We have also investigated the positions of minima in the dispersion curve for magnetorotons, at quantized Hall states of the form $\nu = p / (2p+1)$, in the limit of large $p$, in the absence of impurity scattering.   The minima of interest to us occur at wave vectors $q_n$ that are small compared to $k_F$, and at frequencies that are small compared to the energy gap $\Delta \omega_c = |\Delta B|/m^*$, where $\Delta B$ is the deviation of the magnetic field $B$ from the value corresponding to $\nu=1/2$, at the given electron density.  Therefore, the positions of these minima are properly a subject for investigation in a theory that is supposed to be valid in the limit of long wavelengths and low frequencies. We have found that the HLR and Dirac theories give identical values for the location of these minima, consistent with PH symmetry, at least to order $| \Delta B|^2$.

It is more difficult to compare predictions of the two theories for correlation functions or response functions at a wave vector $q$ that is not small compared to $k_F$, even if the frequency is arbitrarily small. An important example is the correlation function studied by Geraedts et al.\cite{geraedtsnum}. The authors introduce an operator $P(\bfr)$ which is proportional to $n_{\rm{el}} (\bfr)  \nabla^2 n_{\rm{el}} (\bfr)$, projected to the  lowest Landau level, and they study the correlation  function for the Fourier transform, $\langle P_{ - \bfq} P_{\bfq} \rangle$,  for $q$ close to $2 k_F$. According to the Dirac theory, this correlation function should have no observable singularity at $q=2k_F$, because $P(\bfr)$ is even under PH inversion, and fluctuations in such quantifies should not give rise to backscattering across the Fermi surface at $q=2 k_F$. Geraedts et al. have studied this correlation function numerically, for electrons confined to the lowest Landau level at half filling, using density-matrix renormalization group (DMRG) methods, and have found the singularity to be missing, as predicted. By contrast, they do observe a singularity at $q = 2 k_F$, as expected,  in the density correlation function   $\langle  n^{\rm{el}} _{ - \bfq}  n ^{\rm{el}}_{\bfq} \rangle$.

There does not seem to be any  obvious reason in HLR theory  why $\langle P_{ - \bfq} P_{\bfq} \rangle$ should be immune from a singularity at $q=2k_F$, even if one imposes the requirement of particle-hole symmetry. However,
in order to actually calculate this response function in the HLR theory, one would have to know the correct form of the renormalized vertex that couples $P_\bfq$ to the composite fermions at $q=2 k_F$.  It is certainly possible that this quantity will vanish when $m=0$, but at present, we do not have an argument to that effect.  Thus, we cannot say that HLR and the Dirac theories make identical predictions for this property, but we can say that there is not a necessary  contradiction between the two theories, in so far as the relevant vertices are unknown.

The HLR and Dirac theories can both be extended to describe a situation where the  Fermi surface turns out to be unstable to formation of Cooper pairs, with the result that the actual ground state is an incompressible fractional quantized  Hall state, with an energy gap.  As Son has observed, pairing in the Dirac theory must occur in a channel with even angular momentum, because of the Berry phase associated with the Dirac composite fermions. The three most obvious channels for pairing are then $l=0$, $2$, and $-2$.  The symmetries of the $l=2$ and $l=-2$ state coincide, respectively,  with the of the well-known ``Pfaffian"  and ``anti-Pfaffian" states,  which are related to each other by PH conjugation\cite{mooreMR91,readgrn,levinapf,ssletalapf}.  The Dirac theory predicts that these two states should have identical energies, as is indeed required by PH symmetry, in the limit  where electrons there are confined to a single Landau level, and there are only two-body interactions among them.  Within the HLR theory, the Pfaffian and anti-Pfaffian states would be described by pairing in
the channels $l=1$ and $l=-3$ respectively.  There is no obvious reason, within the theory, why these two states should have the same energy. However, such a coincidence is perfectly compatible with the theory; it means that for a PH symmetric system, the pairing interaction must be  the same in the $l=1$ and $l=-3$ channels.  Pairing in the $l=0$ channel of the Dirac model would lead to a new PH symmetric quantized Hall state, which Son named the PH-Pfaffian.  Such a state would be described in HLR by pairing in the channel $l=-1$. There does not seem to be any  numerical evidence that such a state would actually be the ground state of any quantum Hall system with realistic parameters. Wang and Chakravarty\cite{Chakravarty} argued that within a particular approximation scheme, the $l=0$ pairing appears to be unfavorable in Dirac composite fermi liquid. However, Zucker and Feldman have suggested  that the PH-Pfaffian state seems compatible with existing experiments, and the state could have been stabilized by disorder and Landau-level mixing\cite{feldman}.
(The PH-Pfaffian is equivalent to the ``T-Pfaffian" state, which was proposed, independently, in the context of surface states of topological superconductors\cite{fidkowski3d}.)

In summary, we have found no contradictions between physical predictions of the HLR and Son-Dirac theories for the low-energy properties of a half-filled Landau level. We find that the HLR approach is quite compatible with the existence of particle-hole symmetry, which is required in the case where the bare electron mass is taken to zero.  For some properties this symmetry emerges automatically from the HLR theory, while in other cases it may be necessary to properly specify the value of parameters such as the Landau interactions strengths or a renormalized finite-momentum vertex. These results are all consistent with the point of view that the physics described by the particle-hole symmetric Son-Dirac theory is in fact a special case of the HLR theory.

As this manuscript was nearing completion, however, we became aware of recent work by M.  Levin and D. T. Son, which asserts that the HLR approach is not able to obtain the correct value for the Hall viscosity at $\nu=1/2$, in the PH symmetric limit\cite{LevinSon}.  The Hall viscosity is reflected in a  correction to the Hall conductance at non-zero wavevector q, which  appears in the limit  $q \to 0$  and $ \omega \to 0$, with $\omega \gg q v_F^*$. Although the Hall viscosity may be very difficult to measure experimentally, this suggests that there are theoretical problems that need to be resolved before we can determine the precise relation between the HLR and Son-Dirac theories. {Therefore it is still possible that the two theories may eventually be physically distinct, in which case the difference in their measurable behaviors would be much subtler than previously believed. Of course, even if both theories agree, it remains possible that neither one is correct in all respects.}


\subsection*{Acknowledgments}
The authors acknowledge stimulating discussions with T. Senthil, S. Raghu, D. T. Son, and D. Mross. We thank Raghu and Son  for sending us advanced copies of their respective works. 
CW is supported by the Harvard Society of Fellows. This work was also supported, in part, by the Microsoft Corporation Station Q,  by EPSRC Grant no. EP/J017639/1, by the 
European
Research Council under the European Unions
Seventh Framework Program (FP7/2007-2013) / ERC
Project MUNATOP,  by the DFG (CRC/Transregio 183, EI
519/7-1), by the Minerva Foundation, and by the U.S.-Israel BSF.


\bibliography{DiracHLR}

\end{document}